\documentclass[twocolumn,showpacs,preprintnumbers,amsmath,amssymb,prb]{revtex4}
\usepackage{graphicx}% Include figure files
\usepackage{dcolumn}% Align table columns on decimal point
\usepackage{bm}% bold math

\begin{document}

%\preprint{APS/123-QED}

\title{Extended variational cluster approximation for correlated systems}  % Force line breaks with \\

\author{Ning-Hua Tong}
 \address{
Institut f\"{u}r Theorie der Kondensierten Materie \\
          Universit\"{a}t Karlsruhe, 76128 Karlsruhe, Germany \\}

%\author{Second Author}%
% \email{Second.Author@institution.edu}
%\affiliation{%
%Authors' institution and/or address\\
%This line break forced with \textbackslash\textbackslash
%}%

%\author{Charlie Author}
% \homepage{http://www.Second.institution.edu/~Charlie.Author}
%\affiliation{
%Second institution and/or address\\
%This line break forced% with \\
%}%

\date{\today}% It is always \today, today,
             %  but any date may be explicitly specified

\begin{abstract}
The variational cluster approximation (VCA) proposed by M. Potthoff {\it et al.} [Phys. Rev. Lett. {\bf 91}, 206402 (2003)]
 is extended to electron or spin systems with
nonlocal interactions. By introducing more than one source field in the action and employing the Legendre transformation,
 we derive a generalized self-energy functional with stationary properties.
 Applying this functional to a proper reference system, we construct the extended VCA (EVCA).
 In the limit of continuous degrees of freedom for the reference system,
EVCA can recover the cluster extension of the extended dynamical mean-field theory (EDMFT).
For a system with correlated hopping, the EVCA recovers the cluster extension of the
dynamical mean-field theory for correlated hopping.
Using a discrete reference system composed of decoupled three-site single impurities,
we test the theory for the extended Hubbard model.
 Quantitatively good results as compared with EDMFT are obtained.
  We also propose VCA (EVCA) based on clusters with periodic boundary conditions.
  It has the (extended) dynamical cluster approximation
  as the continuous limit. A number of related issues are discussed.

\end{abstract}

% PACS, the Physics and Astronomy
                             % Classification Scheme.
%\keywords{Suggested keywords}%Use showkeys class option if keyword
                              %display desired
\maketitle

\section{Introduction}

In the past decade, the dynamical mean-field theory (DMFT) was developed as a powerful technique for the study of
correlated electron systems. \cite{Metzner1,Georges1} Much knowledge have been obtained
on the strong-correlation effect in three-dimensional systems,
 such as the Mott-Hubbard metal-insulator transition,
\cite{Georges1,Bulla1,Tong1} itinerant ferromagnetism, \cite{Wahle1,Vollhardt1} and the colossal magnetic
resistance in manganites,\cite{Millis1} etc. Recently there are DMFT studies combined with
{\it ab initio} energy band calculations for
 real materials.\cite{Held1,Vollhardt2} One of the many ways to derive the dynamical mean-field theory is
to approximate the exact Luttinger-Ward functional by a local one which depends only on the local Green's functions.
 \cite{Georges1}
The latter is then obtained from an effective single-impurity model with a self-consistently ascribed bath.
This approximation becomes exact in the limit of infinite spatial dimensions. \cite{Metzner1}
For finite-dimensional systems, this approximation ignores the spatial fluctuation effects from the outset,
and  the nonlocal interactions in the model are treated only on the Hartree level.

In order to take into account the spatial fluctuations beyond DMFT, besides the systematic $1/D$ expansion
approach, \cite{Schiller1,Zarand1} various cluster algorithms,
e.g., the cellular DMFT (CDMFT),\cite{Kotliar1,Biroli1,Biroli2,Bolech1} and the dynamical cluster approximation (DCA),
 \cite{Hettler1} have been proposed. The former uses clusters with open boundary conditions, and
 the latter has periodic boundary conditions. These methods not only produce a ${\bf k}$-dependent self-energy,
 but also include the short-range part of the nonlocal interaction within the cluster.
 On the other hand, DMFT has also been extended
 to treat the nonlocal interaction terms in the Hamiltonian more satisfactorily. In this respect, there is
  the extended DMFT (EDMFT) for including the nonlocal density-density
 or spin-spin interactions,\cite{Si1,Kajueter1,Sun1} and the DMFT for correlated hopping\cite{Schiller2,Shvaika1}
  (DMFTCH) for treating
 the correlated hopping terms. An approach based on fictive impurity models has been proposed
 to effectively circumvent the possible noncausality problem in previous cluster algorithms.\cite{Okamoto1}
  All these developments have led to various
 dynamical mean-field algorithms and their cluster extensions, \cite{Maier1}
  which constitute one of the most active fields of the strongly correlated electron theories.

Recently, another cluster theory, the variational cluster approximation (VCA),
was proposed by Potthoff {\it et al.}. \cite{Potthoff1}
This theory provides a rather general framework to formulate cluster approximations for a lattice fermion systems.
In the VCA, different approximations are realized by using different reference systems.
Such reference systems can have different degree of complexity, ranging from a two-site system\cite{Potthoff2}
  to a cluster embedded in several auxiliary baths. In the limit of continuous bath degrees of freedom,
  the VCA recovers the formula of CDMFT in the simple basis.
The DMFT is obtained from the VCA by using a single-impurity
reference system with continuous degrees of freedom.

The VCA is constructed from a self-energy functional which is obtained by doing a Legendre transformation on
 the Luttinger-Ward functional $\Phi \left[G \right]$. \cite{Luttinger1,Baym1}
 The Luttinger-Ward functional plays an important role in the theories
of correlated fermion systems. It was first constructed in 1960s in the course of expressing the thermodynamical
 grand potential through perturbative skeleton diagram expansions. \cite{Luttinger1} Later, it was shown that
 this functional is very useful in constructing conserving and consistent approximations for
  correlated many-body systems. \cite{Baym1,Baym2}
The derivation of the VCA crucially depends on the following properties of the
Luttinger-Ward functional. \cite{Potthoff1} (i) The self-energy as a functional of the full Green's function is given
by the functional derivation of $\Phi \left[ G \right]$,
 \begin{equation}
     \Sigma \left[ { G} \right]=\frac{\delta \Phi \left[ G \right] }{\delta G} \, .      \label{1}
  \end{equation}
(ii) When evaluated at the physical value of the Green's functions, $\Phi \left[ G \right]$ is related to the equilibrium
 grand potential $\Omega$ through the relation, \cite{note1}
\begin{equation}
\beta \Omega=  \Phi\left[  G \right] + {\text Tr} \ln \left[ - G \right] - {\text Tr} \left( {\Sigma G }\right)   \,.   \label{2}
\end{equation}
(iii) $\Phi \left[G \right]$ depends only on the interaction part of the Hamiltonian, and does not contain explicit
dependence on the hopping term.
In the derivation of the VCA, (i) and (ii) guarantee that the grand potential as a functional of the
self-energy $\Omega \left[ {\Sigma} \right]$ can be constructed by Legendre transformation.
This functional is stationary at the physical values of the self-energy $\Sigma$.
(iii) guarantees that any reference system sharing the same interacting part of the original $H$ will
have the same form of the functional $\Phi \left[ G \right]$, and hence the same form of the functional
$\Omega \left[ \Sigma \right]$ (although their actual values depend on the hopping part and may well be different).
The VCA is constructed by identifying the stationary point of $\Omega \left[  \Sigma \right]$ in a self-energy function
space limited by the reference systems as the approximate solution of $\Sigma$. One of the merits of the VCA is that by
using a reference system with small decoupled clusters and limiting the number of variational parameters, thermodynamically
consistent and systematic physical results for a correlated electron system can be obtained with much less numerical
effort. \cite{Potthoff1,Dahnken1,Aichhorn1}

One open question is, in the spirit of the VCA, how one can take into account the nonlocal interaction term beyond the
Hartree approximation. In the VCA, it is possible to include longer- and longer-range nonlocal interactions by increasing
the size of the cluster. However, for interactions beyond the cluster size, employing the Hartree approximation
amounts to defining a reference system with a different nonlocal interaction from the original Hamiltonian, and hence
with different functional $\Omega \left[ \Sigma \right]$.
Thus the condition for establishing the VCA is not fulfilled exactly and further approximation beyond the VCA
is introduced.
In this respect, the EDMFT was devised to incorporate the effect of the nonlocal interaction.
The problems with EDMFT are that first, calculations are numerically very demanding because of the introduction of
additional continuous bosonic bath(s) in EDMFT. Usually the quantum Monte Carlo (QMC) technique has to be used to
solve the impurity problem, and hence the study is limited to finite temperatures.\cite{Sun1}
Second, the EDMFT is basically a single-impurity approach and can produce only local self-energies.
 To obtain the ${\bf k}$-dependent self-energy, one needs a cluster extension of EDMFT such as
the EDMFT$+$CDMFT scheme proposed by Sun {\it et al.}, \cite{Sun1} or the extended version of the
DCA for including nonlocal interactions.\cite{Hettler1} Both of them are very heavy numerical tasks.
  Therefore, it is desirable to extend the VCA to handle the systems with nonlocal interactions. In the continuous limit,
 the EDMFT and EDMFT$+$CDMFT should be recovered if one takes a single-impurity reference system and
 cluster reference system, respectively. Hopefully the flexibility and efficiency of the VCA will enhance the study of correlated electrons
  with nonlocal interactions. Within the same framework, this kind of extended the VCA (EVCA) can be
  constructed for a system
 with correlated hoppings. For this system, the EVCA theory in the continuous limit should recover the DMFTCH
 if one uses a single-impurity reference system, and become a cluster extension of DMFTCH if one uses a cluster
 reference system.

The purpose of this paper is to present such an extension of the VCA. The original Luttinger-Ward functional is a functional
 of the full one-particle Green's function. To extend the VCA,  we first express the grand potential
 functional $\Omega \left[ G, \Pi\right]$ by the generalized Luttinger-Ward functional
 $\Phi \left[ G, \Pi\right]$.\cite{Chitra1} This is accomplished through a Legendre transformation.\cite{Fukuda1}
 Here, besides the one-particle
 Green's function $G$, a certain two particle-Green's function $\Pi$ should be chosen according to the
  form of the nonlocal interactions. For the construction of $\Phi \left[ G, \Pi\right]$,
 we use a derivation that is formally different from that of Chitra {\it et al.},\cite{Chitra1}
 but is more suitable for the present purpose. This is embodied, e.g.,
 in that the universality of $\Phi \left[G, \Pi \right]$ and $\Omega \left[G, \Pi \right]$
 is easily understood from the derivation process. Recently such a nonperturbative construction has appeared for
  the Luttinger-Ward functional $\Phi \left[G \right]$. \cite{Potthoff2} In our work a similar construction
  for the case of multiple variables is carried out.

 Starting from $\Omega \left( G, \Pi\right)$, the generalized self-energy
 functional \cite{Potthoff1} $\Omega_{EVCA} \left[ \Sigma, \Gamma \right]$
is obtained by doing another Legendre transformation with respect to $G$ and $\Pi$. The stationary value
of this functional equals the exact grand potential $\Omega$ of the original system,\cite{note2} and it is reached
at the exact value of the self-energies. In contrast to $\Omega \left[ \Sigma \right]$ which is only independent of hopping term
of the Hamiltonian, the form of $\Omega_{EVCA} \left[ \Sigma, \Gamma \right]$ is independent of not only the hopping term,
 but also the nonlocal interactions. It is hence possible to evaluate it in a clusterized
reference system sharing local interactions with the original Hamiltonian.
Finally, the EVCA is obtained by identifying the stationary point in the generalized self-energy space of a reference system
as the approximate solution of the original system.
This part of our construction is in the same spirit as the VCA, \cite{Potthoff1} although
  the derivation is formally different.

In this paper, as specific examples,
we describe the extension of the VCA for two systems with nonlocal interactions.
One is the Hubbard model with nonlocal density-density
Coulomb repulsion, the other is the Hubbard model with correlated hopping. In both cases, the Hamiltonians have a
nonlocal interaction term that can be treated on the Hartree level within DMFT, and can be partly taken into account
within the cluster
in CDMFT or DCA methods. For the Hubbard model with density-density interaction, numerical results are given for
the simplest reference system, in which each decoupled cluster is a single impurity problem with three sites.
 We will call this scheme three-site EVCA.
Its results are found to be quantitatively close to the EDMFT results.
Finally, based on the real space formula of the DCA by Biroli {\it et al.}, \cite{Biroli1}
 we propose a VCA and EVCA that utilize clusters with periodic boundary conditions.
 In the continuous limit, this version of the VCA will lead to the DCA instead of the CDMFT,
 and correspondingly, the EVCA formula will lead to the extended version of DCA.

 The plan of this paper is the following. In Sec. II, we present the construction of the EVCA for the Hubbard model
 with nonlocal density-density interaction. In Sec. III, the EVCA is constructed for the Hubbard model with correlated
 hopping. Section IV is devoted to the EVCA formula with translation invariance. We discuss some issues related
 to the VCA and EVCA in Sec. V. A summary is given in Sec. VI.

\section{Nonlocal Density-Density Interaction}

In this section, we will first obtain a grand potential functional $\Omega \left[ G, \Pi \right]$ expressed via
the generalized Luttinger-Ward functional $\Phi \left[ G, \Pi \right]$. The grand potential functional
$\Omega_{LW} \left[ G, \Pi \right]$ that bears stationary properties is obtained as a byproduct.
The EVCA formula is established in general form.
Numerical calculations are carried out for the three-site EVCA and results are tested.

 Our construction of $\Phi\left[ G, \Pi \right]$ is nonperturbative and
 does not employ the diagrammatic expansion. Essentially,
 the way that we construct the multiple variable Luttinger-Ward functional is the same as that used by
 Baym \cite{Baym2} for one variable case,
but here a Legendre transformation is employed explicitly.  Therefore, our derivation relies neither on the
applicability of the Wick's theorem, nor on the convergence of the summation of the skeleton diagrams.
It only relies on the assumption that the Legendre transformation,
whenever required, can be carried out. This assumption has been argued to be correct for the one-variable
case. \cite{Potthoff3}

Besides interacting one-particle Green's function, the generalized Luttinger-Ward functional is supposed to
be also a functional of two-particle Green's functions. There are various two-particle Green's functions, such as
the density-density Green's function, spin-spin Green's function, and the occupation-correlated electron
Green's function, etc. Our way of constructing the generalized Luttinger-Ward functional
is general and does not depend on the type of the Green's function.
The type of the two-particle Green's function is
determined by the specific form of the nonlocal interaction in the system.
 For the Hubbard model with a nonlocal density-density interaction,
 the generalized Luttinger-Ward functional $\Phi$ is considered as a functional of the full Green's function $G$
 and the full density-density Green's function $\Pi_{ij} \left( \tau - \tau^{\prime} \right)= \langle T_{\tau} n_{i} \left( \tau \right)
  n_{j} \left( \tau^{\prime}\right) \rangle$. For other systems, it is straightforward to incorporate
  the specific type of two-particle Green's functions as variables of $\Phi$. The Hubbard model with correlated hopping
  will be considered in Sec.III.

 \subsection{Generalized Luttinger-Ward functional}
We start from the Hubbard model with nonlocal Coulomb repulsion,

\begin{eqnarray}
H & = & -\sum\limits_{i,j, \sigma } t_{i j}c_{i\sigma }^{\dagger}c_{j\sigma } + \sum\limits_{ i,j }V_{i j} n_{i}n_{j}-\mu \sum_{i}n_{i} \nonumber \\
     & &  + H_{loc}  .          \label{3}
\end{eqnarray}
Here, the notations are standard. $\sum_{i,j}$ indicates the sum of sites $i$ and $j$ independently.
$\mu$ is the chemical potential, and $H_{loc}= U \sum_{i} n_{i \uparrow} n_{i \downarrow} $ is the local Hubbard interaction.
This model was recently studied using EDMFT.\cite{Chitra2}
The statistical action for Eq. (3) reads

\begin{eqnarray}
  & & S [ c^{*}, c, G_{0}, \Pi_{0} ]        \nonumber \\
   &=&  \int_{0}^{\beta} d \tau \int_{0}^{\beta} d \tau^{\prime} \sum_{i j \sigma} c_{i \sigma}^{*} \left(\tau \right)
          [ - G_{0}^{-1} ]_{i j \sigma} \left( \tau - \tau^{\prime} \right) c_{j \sigma} \left( \tau^{\prime} \right)          \nonumber \\
   &+&   \int_{0}^{\beta} d \tau \int_{0}^{\beta} d \tau^{\prime} \sum_{i j} n_{i} \left(\tau \right)
          [ - \Pi_{0}^{-1} ]_{i j} \left( \tau - \tau^{\prime} \right) n_{j} \left( \tau^{\prime} \right)            \nonumber \\
   &+&  \int_{0}^{\beta} H_{loc} \left(\tau \right) d \tau              ,                 \label{4}
\end{eqnarray}
where

\begin{equation}
-G_{0 ij}^{-1} \left( \tau -\tau^{\prime}\right) = \left[ \left( \frac{\partial}{\partial \tau} - \mu\right) \delta_{i j} -t_{i j} \right]
                        \delta \left(\tau - \tau^{\prime} \right)                   \label{5}
\end{equation}
and
\begin{equation}
-\Pi_{0 i j}^{-1} \left( \tau -\tau^{\prime}\right) = V_{i j} \delta \left( \tau - \tau^{\prime} \right)           .  \label{6}
\end{equation}

The grand partition function $\Xi$ is expressed through the path integral as

\begin{equation}
     \Xi \left[ G_{0}, \Pi_{0} \right] = \int \prod_{i \sigma}  {\mathcal D} c^{*}_{i \sigma} {\mathcal D} c_{i \sigma}
                                  e^{-S \left[c^{*}, c, G_{0}, \Pi_{0} \right]}       ,           \label{7}
\end{equation}
and the grand potential of the system is

\begin{equation}
           \Omega \left[ G_{0}, \Pi_{0} \right] = - \frac{1}{\beta} \ln  \Xi \left[ G_{0}, \Pi_{0} \right]   .         \label{8}
\end{equation}
Equation (8) in fact gives out the grand potential $\Omega$ as a functional of $G_{0}$ and $\Pi_{0}$.
In the following, we will do the Legendre transformation on the functional $\Omega \left[ G_{0}, \Pi_{0}\right]$.
However, since we have explicitly assigned values to $G_{0}$ and $\Pi_{0}$ in Eqs. (5) and (6),
we introduce $\widetilde{G}_{0}$ and $\widetilde{\Pi}_{0}$ as variables of the
grand potential functional. For the clarity of the derivation, it is useful
to distinguish a variational variable and the physical value of it.
In the following, we use the symbol with a tilde on top to denote a quantity
that can be varied, and the symbol without a tilde to denote the fixed physical value of that quantity.

$\widetilde{\Omega}$ as a functional of $\widetilde{G}_{0}$ and $\widetilde{\Pi}_{0}$ is then defined as
\begin{equation}
     \widetilde{ \Omega} \left[ \widetilde{G}_{0}, \widetilde{ \Pi} _{0} \right] = - \frac{1}{\beta} \ln  \widetilde{\Xi}
      \left[ \widetilde{G}_{0}, \widetilde{\Pi}_{0} \right]                          \,\, .                   \label{9}
\end{equation}
Similar to Eqs. (7) and (4), we define $\widetilde{\Xi}$ and  $\widetilde{S}$ as

\begin{equation}
    \widetilde{\Xi} \left[ \widetilde{G}_{0}, \widetilde{\Pi}_{0} \right] = \int \prod_{i \sigma}
    {\mathcal D} c^{*}_{i \sigma} {\mathcal D} c_{i \sigma}
    e^{-\widetilde{S} \left[c^{*}, c,\widetilde{G}_{0}, \widetilde{\Pi}_{0} \right]}            \label{10}
\end{equation}
and
\begin{eqnarray}
  & & \widetilde{S} [ c^{*}, c, \widetilde{G}_{0}, \widetilde{\Pi}_{0} ]        \nonumber \\
   &=&  \int_{0}^{\beta} d \tau \int_{0}^{\beta} d \tau^{\prime} \sum_{i j \sigma} c_{i \sigma}^{*} \left(\tau \right)
          [ - \widetilde{G}_{0}^{-1} ]_{i j \sigma} \left( \tau - \tau^{\prime} \right) c_{j \sigma} \left( \tau^{\prime} \right)          \nonumber \\
   &+&   \int_{0}^{\beta} d \tau \int_{0}^{\beta} d \tau^{\prime} \sum_{i j} : n_{i} \left(\tau \right) :
          [ - \widetilde{\Pi}_{0}^{-1} ]_{i j} \left( \tau - \tau^{\prime} \right) : n_{j} \left( \tau^{\prime} \right) :           \nonumber \\
   &+&  \int_{0}^{\beta} H^{\prime}_{loc} \left(\tau \right) d \tau.                              \label{11}
\end{eqnarray}
In the above equation, we have singled out the Hartree contribution from the interaction and put it into the local
part $H^{\prime}_{loc}$. We have  $:n_{i} \left( \tau \right): = n_{i} \left( \tau \right) - \langle n_{i} \rangle $ and
\begin{equation}
H^{\prime}_{loc}\left( \tau \right) =
H_{loc}\left( \tau \right)+ \sum_{i} p_{i} \left[ 2n_{i} \left( \tau \right) - \langle n_{i} \rangle \right]   ,      \label{12}
\end{equation}
where $p_{i}=\sum_{j} \int_{0}^{\beta} d\tau^{\prime}[-\Pi_{0}^{-1}]_{ij}\left( \tau -\tau^{\prime}\right) \langle n_j \rangle$.
Note that the $[-\Pi_{0}^{-1}]$ in $p_i$ is used here as a fixed value Eq. (6).
By definition, $\widetilde{\Omega}$ equals the physical value $\Omega$ when $\widetilde{G}_{0}= G_{0}$ and
$\widetilde{\Pi}_{0}= \Pi_{0}$. From Eq. (11), it is important to observe that the functional form of
$\widetilde{\Omega} [ \widetilde{G}_{0},\widetilde{\Pi}_{0} ]$ is determined by the form of $H^{\prime}_{loc}$ as
well as the specific way of introducing $\widetilde{G}_{0}$ and $\widetilde{\Pi}_{0}$ in Eq. (11).
It does not depend on the values of $\widetilde{G}_{0}$ and $\widetilde{\Pi}_{0}$.

We define the interacting one-particle Green's function $\widetilde{G}$ and the connected two-particle density-density
Green's function $\widetilde{\Pi}$ as

\begin{eqnarray}
     & & \widetilde{G}_{ij \alpha} \left(\tau - \tau^{\prime} \right)                                   \nonumber \\
     &=&  - \langle T_{\tau} c_{i \alpha} \left(\tau \right)
              c_{j \alpha}^{\dagger} \left(\tau^{\prime} \right)\rangle_{\widetilde{S}}   \nonumber \\
     &=& \frac{-1}{\widetilde{\Xi} \left[ \widetilde{G}_{0}, \widetilde{\Pi}_{0} \right]}
	      \int \prod_{i \sigma}  {\mathcal D} c^{*}_{i \sigma} {\mathcal D} c_{i \sigma}
                c_{i \alpha} \left(\tau \right) c_{j \alpha}^{*} \left( \tau^{\prime} \right)   \nonumber \\
    && 	\,\,\,\,      \times e^{-\widetilde{S} \left[c^{*}, c,\widetilde{ G}_{0}, \widetilde{\Pi}_{0} \right]} ,     \label{13}
\end{eqnarray}

\begin{eqnarray}
     & &  \widetilde{\Pi}_{ij} \left(\tau - \tau^{\prime} \right)                    \nonumber \\
      &=&  \langle T_{\tau} \left[ n_{i} \left(\tau \right) -\langle n_{i} \rangle \right]
                 \left[ n_{j } \left(\tau^{\prime} \right) - \langle n_{j} \rangle \right] \rangle_{\widetilde{S}}   \nonumber \\
       &=& \frac{1}{\widetilde{\Xi} \left[ \widetilde{G}_{0}, \widetilde{\Pi}_{0} \right]}
		        \int  \prod_{i \sigma}  {\mathcal D} c^{*}_{i \sigma} {\mathcal D} c_{i \sigma}  \nonumber \\
  &&             \times \left[ n_{i} \left(\tau \right) - \langle n_{i}\rangle \right]
                       \left[ n_{j } \left( \tau^{\prime} \right)-\langle n_{j} \rangle \right]
   	  e^{-\widetilde{S} \left[c^{*}, c,\widetilde{G}_{0}, \widetilde{\Pi}_{0} \right]}	    \, .     \label{14}
\end{eqnarray}

Note that $\widetilde{G}\left(\tau - \tau^{\prime} \right)$ is anti-periodic in $\left(\tau - \tau^{\prime} \right)$
and $\widetilde{\Pi}\left(\tau - \tau^{\prime} \right)$ is periodic, with the period $\beta$. The Fourier transformation
of them corresponds to
functions of fermionic and bosonic Matsubara frequencies, respectively.
 With the above definitions, it is easily seen that

\begin{equation}
        \widetilde{G}_{ij \alpha} \left(\tau - \tau^{\prime} \right) = -\beta
	         \frac{\delta \widetilde{\Omega} \left[ \widetilde{G}_{0},\widetilde{\Pi}_{0} \right]}
		 {\delta \left[ \widetilde{G}_{0}^{-1} \right]_{j i \alpha} \left(\tau^{\prime} - \tau\right)  }          \label{15}
\end{equation}
and
\begin{equation}
        \widetilde{\Pi}_{ij} \left(\tau - \tau^{\prime} \right) = -\beta
	         \frac{\delta \widetilde{\Omega} \left[ \widetilde{G}_{0},\widetilde{\Pi}_{0} \right]}
		 {\delta \left[ \widetilde{\Pi}_{0}^{-1} \right]_{j i} \left(\tau^{\prime} - \tau\right)  }        .  \label{16}
\end{equation}
In the following, we will abbreviate them as
\begin{equation}
        \widetilde{G} \left[ \widetilde{G}_{0}, \widetilde{\Pi}_{0} \right] =- \beta
	                \frac{\delta \widetilde{\Omega} \left[ \widetilde{G}_{0},\widetilde{\Pi}_{0} \right]}
			{\delta \widetilde{G}_{0}^{-1}}                                \label{17}
\end{equation}
and
\begin{equation}
        \widetilde{\Pi} \left[ \widetilde{G}_{0}, \widetilde{\Pi}_{0} \right] =- \beta
	                \frac{\delta \widetilde{\Omega} \left[ \widetilde{G}_{0},\widetilde{\Pi}_{0} \right]}
			{\delta \widetilde{\Pi}_{0}^{-1}}           .                \label{18}
\end{equation}
As functionals of $\widetilde{G}_{0}$ and $\widetilde{\Pi}_{0}$, $\widetilde{G}$ and $\widetilde{\Pi}$
will get their respective
physical values when $\widetilde{G}_{0}= G_{0}$ and $\widetilde{\Pi}_{0}=\Pi_{0}$. Note that for nonzero
 $\widetilde{G}_{0}^{-1}$ and $\widetilde{\Pi}_{0}^{-1}$, even in the case $H_{loc}=0$, the
 calculation of $\widetilde{G}$ and $\widetilde{\Pi}$ is a complicated many-particle problem and
  their values are in general not equal to $\widetilde{G}_{0}$ and $\widetilde{\Pi}_{0}$.

To do the double Legendre transformation for $\widetilde{\Omega} [ \widetilde{G}_{0}, \widetilde{\Pi}_{0} ]$,
we introduce an intermediate functional $\widetilde{F}  [\widetilde{G}, \widetilde{\Pi} ]$,

\begin{eqnarray}
         \widetilde{F} \left[\widetilde{G}, \widetilde{\Pi} \right] &=& \widetilde{\Omega}
	 \left[\widetilde{G}_{0}, \widetilde{\Pi}_{0} \right]  % \nonumber \\
	        - {\text Tr} \left[ \frac{\delta \widetilde{\Omega} \left[ \widetilde{G}_{0},\widetilde{\Pi}_{0} \right]}
			{\delta \widetilde{G}_{0}^{-1}}  \widetilde{G}_{0}^{-1} \right]             \nonumber \\
	       && - {\text Tr} \left[ \frac{\delta \widetilde{\Omega} \left[ \widetilde{G}_{0},\widetilde{\Pi}_{0} \right]}
			{\delta \widetilde{\Pi}_{0}^{-1}}  \widetilde{\Pi}_{0}^{-1} \right]                 \nonumber \\
	       &=& 	\widetilde{\Omega} \left[\widetilde{G}_{0}, \widetilde{\Pi}_{0} \right]+ \frac{1}{\beta}
	                   {\text Tr} \left( \widetilde{G} \widetilde{G}_{0}^{-1} \right)          \nonumber \\
             &&	   + \frac{1}{\beta} {\text Tr} \left( \widetilde{\Pi} \widetilde{\Pi}_{0}^{-1} \right)	             \, \, .                \label{19}
\end{eqnarray}
Here, the trace ${\text Tr} A= \sum_{k, n, \sigma } A_{k \sigma} \left( i \omega_{n} \right) e^{i \omega_{n} 0^{+}}$
is carried out over space, time, and spin coordinates.
The derivative of $\widetilde{F}$ is obtained as

\begin{equation}
         \frac{\delta \widetilde{F} \left[\widetilde{G}, \widetilde{\Pi} \right]}{\delta \widetilde{G}} =
	             \frac{1}{\beta} \widetilde{G}_{0}^{-1}\left[\widetilde{G}, \widetilde{\Pi} \right]          \label{20}
\end{equation}
and
\begin{equation}
         \frac{\delta \widetilde{F} \left[\widetilde{G}, \widetilde{\Pi} \right]}{\delta \widetilde{\Pi}} =
	             \frac{1}{\beta} \widetilde{\Pi}_{0}^{-1}\left[\widetilde{G}, \widetilde{\Pi} \right]         .          \label{21}
\end{equation}
On the right hand of Eqs. (20)  and (21), we have explicitly written $\widetilde{G}_{0}$ and $ \widetilde{\Pi}_{0}$ as functionals
of $\widetilde{G}$ and $\widetilde{\Pi}$ which are regarded as independent variables.
 That this can be done is implied by our original assumption. From Eq. (19), the functional
 dependence of $\widetilde{\Omega}$ on the one- and two-particle Green's functions is obtained as

\begin{equation}
  \widetilde{\Omega}\left[\widetilde{G}, \widetilde{\Pi} \right]=\widetilde{F} \left[\widetilde{G}, \widetilde{\Pi} \right]- \frac{1}{\beta}
	                   {\text Tr} \left( \widetilde{G} \widetilde{G}_{0}^{-1} \right) - \frac{1}{\beta}
	                   {\text Tr} \left( \widetilde{\Pi} \widetilde{\Pi}_{0}^{-1} \right)	  .              \label{22}
\end{equation}

The next step to construct a generalized Luttinger-Ward functional is to introduce the one-particle
self-energy $\widetilde{\Sigma}$ and the quantity $\widetilde{\Gamma}$. In the following, we will define them as

\begin{equation}
       \widetilde{\Sigma}= \widetilde{G}_{0}^{-1} - \widetilde{G}^{-1}       ,      \label{23}
\end{equation}
\begin{equation}
       \widetilde{\Gamma}= \widetilde{\Pi}_{0}^{-1} + \alpha \widetilde{\Pi}^{-1}       .      \label{24}
\end{equation}
Here, the inverse should be considered as the matrix inverse in the space, time, and spin coordinates.
In the definition of $\widetilde{\Gamma}$, we introduce a real parameter $\alpha$.
Physically, the bosonlike nature of the operator $n_{i}$ may advocate the value $\alpha=1/2$, and it was actually taken
by many authors.\cite{note3} However, since $n_{i}\left( \tau \right)$ is not a strict real boson field, the value of $\alpha$
cannot be determined to be $1/2$ from first principle. We will discuss this issue further in Sec. V B.
Combined with Eqs. (20) and (21), the above definitions lead to the following equations:

\begin{equation}
         \frac{\delta [ \widetilde{F} [\widetilde{G}, \widetilde{\Pi} ] \beta - {\text Tr}
	  \ln \widetilde{G} ]}{ \delta \widetilde{G}}
	   = \widetilde{\Sigma}               \,\, ,     \label{25}
\end{equation}

\begin{equation}
         \frac{\delta [ \widetilde{F} [\widetilde{G}, \widetilde{\Pi} ] \beta +\alpha
	 {\text Tr} \ln \widetilde{\Pi} ]}{ \delta \widetilde{\Pi}}
	   = \widetilde{\Gamma}               \,\, .     \label{26}
\end{equation}
With Eqs. (23)$-$(26), we are now ready to introduce $\widetilde{\Phi} [\widetilde{G}, \widetilde{\Pi} ]$,
 which is a generalization of the original Luttinger-Ward functional $\widetilde{\Phi} [\widetilde{G} ]$.
It is defined as

\begin{equation}
       \widetilde{\Phi} \left[\widetilde{G}, \widetilde{\Pi} \right]= \beta \widetilde{F} \left[\widetilde{G}, \widetilde{\Pi} \right]
                 - {\text Tr} \ln \widetilde{G} + \alpha {\text Tr} \ln \widetilde{\Pi}           .                  \label{27}
\end{equation}

There are several important properties of this functional, parallel to the properties of the original
Luttinger-Ward functional $\Phi \left[ G \right]$.
(i) The functional derivative of $\widetilde{\Phi} [\widetilde{G}, \widetilde{\Pi} ]$ with respect to $\widetilde{G}$ and to
$\widetilde{\Pi}$ gives $\widetilde{\Sigma}[\widetilde{G}, \widetilde{\Pi} ]$ and
$\widetilde{\Gamma}[\widetilde{G}, \widetilde{\Pi} ]$, respectively,
\begin{equation}
        \frac{\delta \widetilde{\Phi} \left[\widetilde{G}, \widetilde{\Pi} \right] }{ \delta \widetilde{G}}
	   = \widetilde{\Sigma}               \,\, ,     \label{28}
\end{equation}

\begin{equation}
        \frac{\delta \widetilde{\Phi} \left[\widetilde{G}, \widetilde{\Pi} \right] }{ \delta \widetilde{\Pi}}
	   = \widetilde{\Gamma}               \,\, .     \label{29}
\end{equation}
This is easily obtained from Eqs. (25)$-$(27). $\widetilde{\Sigma}$ and $\widetilde{\Gamma}$ acquire
 the physical values $\Sigma$ and $\Gamma$ if  they are evaluated at physical Green's functions $\{ G, \Pi \} $.
Here, the physical meaning of $\Gamma= \Pi_{0}^{-1} + \alpha \Pi ^{-1}$
is somewhat obscure. In analogy to the one-particle self-energy, it can be regarded as an effective density
cumulant that plays the role of the self-energy for two-particle Green's functions.\cite{Smith1}
 (ii) The grand potential functional can be expressed by $\widetilde{\Phi}$ as
   \begin{eqnarray}
       \beta \widetilde{\Omega}\left[\widetilde{G}, \widetilde{\Pi} \right] &=&  \widetilde{\Phi}
       \left[\widetilde{G}, \widetilde{\Pi} \right]
	+ {\text Tr} \ln \widetilde{G} - {\text Tr} \left( \widetilde{G} \widetilde{G}_{0}^{-1} \right)    \nonumber \\
	&-& \alpha {\text Tr}  \ln \widetilde{\Pi} - {\text Tr} \left( \widetilde{\Pi} \widetilde{\Pi}_{0}^{-1} \right)     .      \label{30}
 \end{eqnarray}
Here, $\widetilde{G}_{0}^{-1} $ and $\widetilde{\Pi}_{0}^{-1}$ are both considered as functionals of $\widetilde{G}$
and $\widetilde{\Pi}$. Eq. (30) is obtained by combining Eqs. (22) and (27).
Unlike the original Luttinger-Ward functional $\Phi \left[ G \right]$,
$\widetilde{\Phi} [\widetilde{G}, \widetilde{\Pi} ]$ is nonzero in the case of $H_{loc}=0$. Therefore the
explicit expression of
$\widetilde{\Omega}[\widetilde{G}, \widetilde{\Pi} ] $ in this limit is still a nontrivial many-body problem.
(iii) The form of the functional $\widetilde{\Phi} [\widetilde{G}, \widetilde{\Pi} ]$ is determined only by the form of
 $H_{loc}$ once the type of the two-particle Green's function $\widetilde{\Pi}$ is chosen. This is easily understood as
 a consequence of our derivation process. As stated above, the functional dependence of
 $\widetilde{\Omega} [\widetilde{G}_{0}, \widetilde{\Pi}_{0} ]$ on $\{ \widetilde{G}_{0}^{-1},
{\widetilde{\Pi}_{0}^{-1}} \}$ [Eqs. (9)$-$(11)] is determined only by the form of
$H_{loc}$ as well as by the specific form of Eq. (11) in which $\widetilde{G}_{0}^{-1}$ and $\widetilde{\Pi}_{0}^{-1}$ are
introduced.
This is also true for the functional derivatives of $\widetilde{\Omega} [\widetilde{G}_{0}, \widetilde{\Pi}_{0} ]$,
$ \widetilde{G}$ and  $\widetilde{\Pi}$, and their inverse functionals. Following Eq. (19), $\widetilde{F}
[ \widetilde{G},\widetilde{ \Pi} ]$,
the Legendre transformation of $\widetilde{\Omega} [\widetilde{G}_{0}^{-1}, \widetilde{\Pi}_{0}^{-1} ]$ also
bears the same property. And our conclusion follows Eq. (27) immediately.

In the original paper of Luttinger and Ward, a functional expression of
$\widetilde{\Omega}$ that is stationary at the physical value of $\widetilde{G}$ is obtained.
It is very important for devising various kinds of thermodynamically consistent
approximations. Here, the functional form of the grand potential functional that is stationary
at the physical value of $\{ \widetilde{G}, \widetilde{\Pi} \}$ is obtained from a modification on
  $\widetilde{\Omega}[\widetilde{G} ,\widetilde{\Pi} ]$ of Eq. (30); namely, in Eq. (30),
  we replace the $\widetilde{G}_{0}$ and $\widetilde{\Pi}_{0}$ with their respective physical
  values $G_{0}$ and $\Pi_{0}$ defined in Eqs. (5) and (6). We denote the resulting functional
  as $\widetilde{\Omega}_{LW}$. It reads
 \begin{eqnarray}
       \beta \widetilde{\Omega}_{LW}\left[\widetilde{G}, \widetilde{\Pi} \right]
       &=&  \widetilde{\Phi} \left[\widetilde{G}, \widetilde{\Pi} \right]
	+ {\text Tr} \ln \widetilde{G} - {\text Tr} \left( \widetilde{G} G_{0}^{-1} \right)    \nonumber \\
	&-& \alpha {\text Tr}  \ln \widetilde{\Pi} - {\text Tr} \left( \widetilde{\Pi} \Pi_{0}^{-1} \right)     .      \label{31}
 \end{eqnarray}
The functional $\widetilde{\Omega}_{LW} [\widetilde{G}, \widetilde{\Pi} ]$ satisfies the stationary properties
\begin{eqnarray}
  && \frac{\delta \widetilde{\Omega}_{LW} \left[\widetilde{G},\widetilde{\Pi} \right]}
  {\delta \widetilde{G}} |_{\widetilde{G}=G, \widetilde{\Pi}=\Pi}=0  , \nonumber \\
&& \frac{\delta \widetilde{\Omega}_{LW} \left[\widetilde{G},\widetilde{\Pi} \right]}
{\delta \widetilde{\Pi}}|_ {\widetilde{G}=G, \widetilde{\Pi}=\Pi}=0  ,\nonumber \\
&&\widetilde{\Omega}_{LW} \left[\widetilde{G}=G,\widetilde{\Pi}=\Pi \right]= \Omega .   \label{32}
\end{eqnarray}
The above derivation, when reduced to the case of one variable $\widetilde{G}$,
 will lead to the functional $\Omega_{LW} [ \widetilde{G}]$.
 It is actually what Luttinger and Ward obtained by skeleton diagram expansion,
 and can be taken as a starting point of constructing consistent approximations.
However, for the construction of the EVCA in the following, we will not make use of the
form of $\widetilde{\Omega}_{LW} [\widetilde{G}, \widetilde{\Pi} ]$. Instead,
we take the functional form $\widetilde{\Omega} [\widetilde{G}, \widetilde{\Pi} ]$
as a further starting point toward EVCA.

\subsection{EVCA for density-density interaction}

In this section, we continue to derive the EVCA for the Hubbard model with
nonlocal density-density interaction (3). The derivation is based on
the generalized Luttinger-Ward functional $\widetilde{\Phi}[\widetilde{G}, \widetilde{\Pi} ]$
obtained  in Sec. II A.
 The grand potention $\widetilde{\Omega}$ has been expressed by $\widetilde{\Phi}$.
 To go further, we follow the original idea of Potthoff \cite{Potthoff1}
 and apply a Legendre transformation to
 $\widetilde{\Phi}[\widetilde{G}, \widetilde{\Pi} ]$,
\begin{eqnarray}
         \widetilde{A} \left[\widetilde{\Sigma}, \widetilde{\Gamma} \right] &=& \widetilde{\Phi}
	 \left[\widetilde{G}, \widetilde{\Pi} \right]
	  - {\text Tr} \left[ \frac{\delta \widetilde{\Phi} \left[ \widetilde{G},\widetilde{\Pi} \right]}
		{\delta \widetilde{G}}  \widetilde{G} \right]             \nonumber \\
	       && - {\text Tr} \left[ \frac{\delta \widetilde{\Phi} \left[ \widetilde{G},\widetilde{\Pi} \right]}
		{\delta \widetilde{\Pi}}  \widetilde{\Pi} \right]                 \nonumber \\
	       &=& 	\widetilde{\Phi} \left[\widetilde{G}, \widetilde{\Pi} \right]
        - {\text Tr} \left( \widetilde{\Sigma} \widetilde{G} \right)
      - {\text Tr} \left( \widetilde{\Gamma} \widetilde{\Pi}\right)	       .           \label{33}
\end{eqnarray}
For the intermediate functional $\widetilde{A}[ \widetilde{\Sigma}, \widetilde{\Gamma} ]$, we have
\begin{eqnarray}
         \frac{\delta \widetilde{A} \left[\widetilde{\Sigma}, \widetilde{\Gamma} \right]}{\delta \widetilde{\Sigma}} &=&
	            -\widetilde{G}\left[\widetilde{\Sigma} , \widetilde{\Gamma} \right]  , \nonumber \\
\frac{\delta \widetilde{A} \left[\widetilde{\Sigma}, \widetilde{\Gamma} \right]}{\delta \widetilde{\Gamma}} &=&
	            -\widetilde{\Pi}\left[\widetilde{\Sigma} , \widetilde{\Gamma} \right]  .     \label{34}
\end{eqnarray}
With $\widetilde{A}[ \widetilde{\Sigma}, \widetilde{\Gamma}]$,
the $\widetilde{\Omega}$ as a functional of $\widetilde{\Sigma}$ and $\widetilde{\Gamma}$ is
obtained,\cite{note2}

\begin{eqnarray}
    && \beta \widetilde{\Omega}\left[\widetilde{\Sigma}, \widetilde{\Gamma}\right]  \nonumber \\
    &=& \widetilde{A} \left[\widetilde{\Sigma},
  \widetilde{\Gamma} \right]-{\text Tr} \ln \left( \widetilde{G}_{0}^{-1} -\widetilde{\Sigma}\right) + \alpha
  {\text Tr} \ln \left( \widetilde{\Pi}_{0}^{-1}- \widetilde{\Gamma} \right)	  .      \nonumber \\
  &&         \label{35}
\end{eqnarray}
Here, $\{ \widetilde{G}_{0}^{-1}, \widetilde{\Pi}_{0}^{-1} \}$ is considered as the functional
of $\{ \widetilde{\Sigma}, \widetilde{\Gamma} \}$.

To construct the  EVCA, however, we need a grand potential functional of
$\widetilde{\Sigma}$ and $\widetilde{\Gamma} $ which reaches its physical value at a stationary point.
The above functional $\widetilde{\Omega}[\widetilde{\Sigma}, \widetilde{\Gamma}]$ doesnot
 satisfy this requirement. Therefore, similar to the construction of
  $\widetilde{\Omega}_{LW} [\widetilde{G}, \widetilde{\Pi}]$, we suggest to use the following
   functional to establish the EVCA,
\begin{eqnarray}
  && \beta \widetilde{\Omega}_{EVCA}\left[\widetilde{\Sigma}, \widetilde{\Gamma}\right] \nonumber \\
  &=& \widetilde{A} \left[\widetilde{\Sigma}, \widetilde{\Gamma} \right]
  -  {\text Tr} \ln \left(G_{0}^{-1} -\widetilde{\Sigma}\right) + \alpha {\text Tr} \ln \left(\Pi_{0}^{-1}- \widetilde{\Gamma} \right) .
       \nonumber \\     &&                   \label{36}
\end{eqnarray}
In Eq. (36), instead of using the variable $\{ \widetilde{G}_{0}, \widetilde{\Pi}_{0}\}$ as in Eq. (35),
their physical values Eqs. (5) and (6) determined by the original Hamiltonian are used.
It is easy to verify that this functional has the desirable stationary properties, i.e.,
\begin{eqnarray}
  && \frac{\delta \widetilde{\Omega}_{EVCA} \left[\widetilde{\Sigma},\widetilde{\Gamma} \right]}
  {\delta \widetilde{\Sigma}} |_{\widetilde{\Sigma}=\Sigma, \widetilde{\Gamma}=\Gamma}=0  , \nonumber \\
&& \frac{\delta \widetilde{\Omega}_{EVCA} \left[\widetilde{\Sigma},\widetilde{\Gamma} \right]}
{\delta \widetilde{\Gamma}}|_{\widetilde{\Sigma}=\Sigma, \widetilde{\Gamma}=\Gamma}=0  ,  \label{37}
\end{eqnarray}
and\cite{note2}
\begin{equation}
\widetilde{\Omega}_{EVCA} \left[\widetilde{\Sigma}=\Sigma,\widetilde{\Gamma}=\Gamma \right]= \Omega
.   \label{38}
\end{equation}
Another important merit of $\widetilde{\Omega}_{EVCA}$ is that its functional dependence
on $\{ \widetilde{\Sigma}, \widetilde{\Gamma} \}$ is only determined by the form of
$H_{loc}$, and it has nothing to do with the concrete values of
$\widetilde{\Sigma}$ and $\widetilde{\Gamma}$, or equivalently
of $\widetilde{G}_{0}$ and $\widetilde{\Pi}_{0}$.
This is the consequence of the Legendre transformation applied to
$\widetilde{\Phi} [\widetilde{G}, \widetilde{\Pi} ]$ which also has such a property.
Combining Eqs. (35) and (36) to eliminate $\widetilde{A}$, we can express
$\widetilde{\Omega}_{EVCA}$ by $\widetilde{\Omega}$,
\begin{eqnarray}
  && \widetilde{\Omega}_{EVCA}\left[\widetilde{\Sigma}, \widetilde{\Gamma}\right] \nonumber \\
  &=& \widetilde{\Omega} \left[\widetilde{\Sigma}, \widetilde{\Gamma} \right]
   +  \frac{1}{\beta} {\text Tr} \ln \left(\widetilde{G}_{0}^{-1} -\widetilde{\Sigma}\right)
    - \alpha \frac{1}{\beta}{\text Tr} \ln \left(\widetilde{\Pi}_{0}^{-1}- \widetilde{\Gamma} \right)   \nonumber \\
  && - \frac{1}{\beta} {\text Tr} \ln \left(G_{0}^{-1} -\widetilde{\Sigma}\right)
      + \alpha \frac{1}{\beta} {\text Tr} \ln \left(\Pi_{0}^{-1}- \widetilde{\Gamma} \right).            \label{39}
\end{eqnarray}
This equation plays a fundamental role in the construction of EVCA.
Based on the same functional for the reference system, we can introduce the EVCA.

To conveniently express the EVCA equations, we first define the reference system,
apply Eq. (39) to the reference system, and finally introduce
the approximation that leads to the EVCA equations.
The reference system is composed of
identical clusters of original lattice sites, and each site in the cluster
may be coupled to a number of noninteracting bath sites.
Hoppings and nonlocal interactions are allowed only inside each cluster, while the local interaction
  $H_{loc}$ on each site is the same as in the original Hamiltonian. For clarity, we
   introduce a subscript $c$ to denote a matrix with zero matrix element between sites on
   different clusters.
    For an example, $\widetilde{G}_{cij}=0$ if sites $i$ and $j$ belong to different clusters.
  Therefore, we can always write matrix $\widetilde{G}_{c}$ in block diagonal form by
   numbering the lattice sites in an appropriate order. $\widetilde{G}_{c}^{I}$ denotes
    the submatrix on the $I$th cluster. To project out the block diagonal part
    of a general matrix $A$, we use the symbol $|A|$. That is,
       $|A|_{ij} = A_{ij}$ if $i$ and $j$ belong to same cluster, and
 $|A|_{ij} =0$ otherwise. The submatrix of $A$
 on the $I$th cluster is denoted by $|A|^{I}$.

The action of a general reference system is written as
\begin{equation}
  \widetilde{S}_{ref} [ c^{*}, c, \widetilde{G}_{0c}, \widetilde{\Pi}_{0c} ]
  = \sum_{I} \widetilde{S}_{c} [ c^{*}, c, \widetilde{G}_{0c}^{I}, \widetilde{\Pi}_{0c}^{I} ]      \label{40}
\end{equation}
and
\begin{eqnarray}
  & & \widetilde{S}_{c} [ c^{*}, c, \widetilde{G}_{0c}^{I}, \widetilde{\Pi}_{0c}^{I} ]        \nonumber \\
   &=&  \int_{0}^{\beta} d \tau \int_{0}^{\beta} d \tau^{\prime} \sum_{ i j \in I \sigma} c_{i \sigma}^{*} \left(\tau \right)
          [ -  \widetilde{G}_{0c} ^{-1} ]^{I}_{i j \sigma} \left( \tau - \tau^{\prime} \right) c_{j \sigma} \left( \tau^{\prime} \right)          \nonumber \\
   &+&   \int_{0}^{\beta} d \tau \int_{0}^{\beta} d \tau^{\prime} \sum_{i j \in I}
              :n_{i}: \left(\tau \right)
          [ - \widetilde{\Pi}_{0c}^{-1} ]^{I}_{i j} \left( \tau - \tau^{\prime} \right)
	  :n_{j}: \left( \tau^{\prime} \right)  \nonumber \\
   &+&  \int_{0}^{\beta} H_{loc}^{\prime I} \left(\tau \right) d \tau                     .                \label{41}
\end{eqnarray}
 Here, $H_{loc}^{ \prime I}= U \sum_{i \in I} n_{i \uparrow} n_{i \downarrow}
  + \sum_{i \in I} p_{i} \left[ 2n_{i} \left( \tau \right) - \langle n_{i} \rangle \right]$,
  $:n_{i}:=n_{i}-\langle n_{i}\rangle$,  and
 $p_{i}=\sum_{j} \int_{0}^{\beta} d\tau^{\prime}[-\Pi_{0}^{-1}]_{ij}\left( \tau -\tau^{\prime}\right) \langle n_j \rangle$.
  Here and in the following, we use
 $\widetilde{S}_{c}$, $\widetilde{\Xi}_{c}$, and $\widetilde{\Omega}_{c}$ to denote the
 corresponding functionals on each cluster. Since the clusters are geometrically identical,
 the functional is of the same form on each cluster, although the values of their variables could be
 $I$ dependent.
The partition function and the grand potential for the reference system are then written as
\begin{eqnarray}
 \widetilde{\Xi}\left[ \widetilde{G}_{0c}, \widetilde{\Pi}_{0c}\right] &=& \prod_{I} \widetilde{\Xi}_{c}
   \left[ \widetilde{G}_{0c}^{I}\right]    \nonumber \\
   &=& \int \prod_{I, i \in I, \sigma}  {\mathcal D} c^{*}_{i \sigma} {\mathcal D} c_{i \sigma}
   e^{-\widetilde{S}_{c}^{I}  [ c^{*}, c, \widetilde{G}_{0c}^{I}, \widetilde{\Pi}_{0c}^{I} ] }      \nonumber\\
    && \label{42}
\end{eqnarray}
and
\begin{eqnarray}
   \widetilde{\Omega} \left[ \widetilde{G}_{0c}, \widetilde{\Pi}_{0c} \right] &=&
    \sum_{I} \widetilde{\Omega}_{c} \left[ \widetilde{G}_{0c}^{I}, \widetilde{\Pi}_{0c}^{I} \right]    \nonumber \\
    &=& -\frac{1}{\beta} \sum_{I} \ln \widetilde{\Xi}_{c} \left[ \widetilde{G}_{0c}^{I},
    \widetilde{\Pi}_{0c}^{I} \right]     . \label{43}
 \end{eqnarray}
The Green's functions are obtained as
\begin{equation}
        \widetilde{G}_{c} \left[ \widetilde{G}_{0c}, \widetilde{\Pi}_{0c} \right] =- \beta
	                \frac{\delta \widetilde{\Omega} \left[ \widetilde{G}_{0c},\widetilde{\Pi}_{0c} \right]}
			{\delta \widetilde{G}_{0c}^{-1}}                            \label{44}
\end{equation}
and
\begin{equation}
        \widetilde{\Pi}_{c} \left[ \widetilde{G}_{0c}, \widetilde{\Pi}_{0c} \right] =- \beta
	                \frac{\delta \widetilde{\Omega} \left[ \widetilde{G}_{0c},\widetilde{\Pi}_{0c} \right]}
			{\delta \widetilde{\Pi}_{0c}^{-1}}             .                \label{45}
\end{equation}

The previous derivations are done for
 arbitrary functions $\widetilde{G}_{0}^{-1}$ and $\widetilde{\Pi}_{0}^{-1}$. They
 also hold for the reference system where  $\widetilde{G}_{0}^{-1}=\widetilde{G}_{0c}^{-1}$
 and $\widetilde{\Pi}_{0}^{-1}=\widetilde{\Pi}_{0c}^{-1}$.
 In particular, the forms of the functionals are not dependent on the variables.
 Therefore, one can use the same set of symbols of the functionals and
 simply replace the variables
 $\widetilde{G}, \widetilde{\Sigma}$, etc.,  with $\widetilde{G}_{c}, \widetilde{\Sigma}_{c}$,
  etc., in each previous expression.
When expressed in the domain of the self-energies of the reference system,
 the functional $\widetilde{\Omega}_{EVCA}[ \widetilde{\Sigma}, \widetilde{\Gamma} $
in Eq. (39) becomes
\begin{eqnarray}
  && \widetilde{\Omega}_{EVCA}\left[\widetilde{\Sigma}_{c}, \widetilde{\Gamma}_{c}\right] \nonumber \\
  &=& \widetilde{\Omega} \left[\widetilde{\Sigma}_{c}, \widetilde{\Gamma}_{c} \right]
   +  \frac{1}{\beta} {\text Tr} \ln \left(\widetilde{G}_{0c}^{-1} -\widetilde{\Sigma}_{c}\right)    \nonumber \\
   &&  - \frac{1}{\beta} {\text Tr} \ln \left(G_{0}^{-1} - \widetilde{\Gamma}_{c} \right)
   - \frac{\alpha}{\beta}{\text Tr} \ln \left(\widetilde{\Pi}_{0c}^{-1} -\widetilde{\Sigma}_{c}\right)  \nonumber \\
   && + \frac{\alpha}{\beta} {\text Tr} \ln \left(\Pi_{0}^{-1}- \widetilde{\Gamma}_{c} \right)   .              \label{46}
\end{eqnarray}

 Since the degrees of freedom in the reference system are decoupled between
 different clusters, the functionals can be expressed as a summation over clusters.
For example, we have
\begin{equation}
     \widetilde{\Omega}\left[ \widetilde{G}_{c}, \widetilde{\Pi}_{c} \right] =
       \sum_{I}\widetilde{\Omega}_{c} \left[ \widetilde{G}_{c}^{I}, \widetilde{\Pi}_{c}^{I} \right]        \label{47}
\end{equation}
and
  \begin{eqnarray}
      && \beta \widetilde{\Omega}_{c}\left[\widetilde{G}_{c}^{I}, \widetilde{\Pi}_{c}^{I} \right]  \nonumber \\
       &=&  \widetilde{\Phi}_{c} \left[\widetilde{G}_{c}^{I}, \widetilde{\Pi} _{c}^{I}\right]
	+ {\text Tr} \ln \widetilde{G}_{c}^{I} - {\text Tr} \left( \widetilde{G} _{c}^{I} [ \widetilde{G}_{0c}^{-1} ]^{I} \right)    \nonumber \\
	&&- \alpha {\text Tr} \ln \widetilde{\Pi}_{c}^{I} - {\text Tr} \left( \widetilde{\Pi} _{c}^{I} [ \widetilde{\Pi}_{0c}^{-1} ]^{I} \right)	.
	 \nonumber \\	&&         \label{48}
 \end{eqnarray}
Here the trace is carried out in the cluster subspace.
To be complete, we write down the intermediate expressions for the reference system.
 \begin{eqnarray}
        \frac{\delta \widetilde{\Phi}_{c} \left[\widetilde{G}_{c}^{I}, \widetilde{\Pi} _{c}^{I}\right] }
	{ \delta \widetilde{G}_{c}^{I}}
	   &=& \widetilde{\Sigma}_{c}^{I} = [\widetilde{G}_{0c}^{-1} ]^{I} - [\widetilde{G}_{c}^{-1} ]^{I} ,             \nonumber \\
        \frac{\delta \widetilde{\Phi}_{c} \left[\widetilde{G}_{c}^{I}, \widetilde{\Pi}_{c}^{I} \right] }{ \delta \widetilde{\Pi}_{c}^{I}}
	   &=& \widetilde{\Gamma} _{c}^{I}
	   =  [\widetilde{\Pi}_{0c}^{-1} ]^{I} + \alpha [\widetilde{\Pi}_{c}^{-1} ]^{I}               .     \label{49}
\end{eqnarray}

Similarly, the generalized self-energy functional $\widetilde{\Omega} [ \widetilde{\Sigma}_{c},
\widetilde{\Gamma}_{c} ]$ reads,
\begin{eqnarray}
     && \beta \widetilde{\Omega}\left[ \widetilde{\Sigma}_{c}, \widetilde{\Gamma}_{c} \right] \nonumber \\
     &=& \beta \sum_{I} \widetilde{\Omega}_{c}
      \left[ \widetilde{\Sigma}_{c}^{I}, \widetilde{\Gamma}_{c}^{I} \right]     \nonumber \\
     &=& \widetilde{A}\left[\widetilde{\Sigma}_{c},  \widetilde{\Gamma}_{c} \right]
     - \sum_{I} {\text Tr} \ln \left( [\widetilde{G}_{0c}^{-1}]^{I} -\widetilde{\Sigma}_{c}^{I}\right)            \nonumber \\
   && + \alpha \sum_{I} {\text Tr} \ln \left( [ \widetilde{\Pi}_{0c}^{-1}]^{I}- \widetilde{\Gamma} _{c}^{I}\right)
      . \label{50}
\end{eqnarray}
The functional $\widetilde{A} [ \widetilde{\Sigma}_{c}, \widetilde{\Gamma}_{c} ]$ is also a summation,
\begin{equation}
   \widetilde{A} \left[ \widetilde{\Sigma}_{c}, \widetilde{\Gamma}_{c} \right]
   = \sum_{I} \widetilde{A}_{c} \left[ \widetilde{\Sigma}_{c}^{I}, \widetilde{\Gamma}_{c}^{I} \right]   ,     \label{51}
  \end{equation}
and $\widetilde{A}_{c}$ satisfies

\begin{eqnarray}
    \frac{\delta \widetilde{A}_{c} \left[\widetilde{\Sigma}_{c}^{I}, \widetilde{\Gamma}_{c}^{I} \right]}
    {\delta \widetilde{\Sigma}_{c}^{I}} &=&
         -\widetilde{G}_{c}^{I}\left[\widetilde{\Sigma}_{c}^{I} , \widetilde{\Gamma}_{c}^{I} \right]  , \nonumber \\
\frac{\delta \widetilde{A}_{c} \left[\widetilde{\Sigma}_{c}^{I}, \widetilde{\Gamma}_{c}^{I} \right]}
{\delta \widetilde{\Gamma}_{c}^{I}} &=&
    -\widetilde{\Pi}_{c}^{I}\left[\widetilde{\Sigma}_{c}^{I} , \widetilde{\Gamma}_{c}^{I} \right]       .     \label{52}
\end{eqnarray}

The EVCA requires that the stationary properties
of $\widetilde{\Omega}_{EVCA}
[ \widetilde{\Sigma}, \widetilde{\Gamma} ]$, Eq. (37) and (38),  be satisfied in
the subspace of $\widetilde{\Sigma}_{c}$ and $\widetilde{\Pi}_{c}$, in
which the functional $\widetilde{\Omega}_{EVCA}
[ \widetilde{\Sigma}_{c}, \widetilde{\Gamma}_{c} ]$ can be exactly obtained by solving
the reference system. At this point the strategy of the EVCA is the same as the VCA.
The EVCA equation is then obtained by
identifying the approximate self-energies of the original
system as the ones at the stationary point of the functional $\widetilde{\Omega}_{EVCA}
[ \widetilde{\Sigma}_{c}, \widetilde{\Gamma}_{c} ]$. It means that we require

\begin{eqnarray}
  && \frac{\delta \widetilde{\Omega}_{EVCA} \left[\widetilde{\Sigma}_{c},\widetilde{\Gamma}_{c} \right]}
  {\delta \widetilde{\Sigma}_{c}}
   |_{\widetilde{\Sigma}_{c}=\Sigma_{app},  \widetilde{\Gamma}_{c}=\Gamma_{app} }=0  ,   \nonumber \\
&& \frac{\delta \widetilde{\Omega}_{EVCA} \left[\widetilde{\Sigma}_{c},\widetilde{\Gamma}_{c} \right]}
{\delta \widetilde{\Gamma}_{c}}
|_{\widetilde{\Sigma}_{c}=\Sigma_{app}, \widetilde{\Gamma}_{c}=\Gamma_{app}} =0        ,       \label{53}
\end{eqnarray}
and\cite{note2}
\begin{equation}
  \Omega_{app}= \widetilde{\Omega}_{EVCA} \left[\widetilde{\Sigma}_{c}=\Sigma_{app},
             \widetilde{\Gamma}_{c}=\Gamma_{app} \right]    ,           \label{54}
\end{equation}
where $\Sigma_{app}$, $\Gamma_{app}$, and $\Omega_{app}$ are the EVCA results for
each quantity. For a reference system with continuous degrees of freedom,
by introducing Eq. (46) and making use of Eqs. (47)$-$(52),
the above equations reduce to
\begin{eqnarray}
  &&   \widetilde{G}_{c}^{I} = \left|  \left( G_{0}^{-1} - \widetilde{\Sigma}_{c}  \right)^{-1} \right|^{I}  \nonumber \\
  &&   \widetilde{\Pi}_{c}^{I} = - \alpha \left|  \left( \Pi_{0}^{-1} - \widetilde{\Gamma}_{c}  \right)^{-1} \right|^{I}  \nonumber \\
  &&  \widetilde{\Sigma}_{c}^{I}= [ \widetilde{G}_{0c}^{-1}]^{I} - [\widetilde{G}_{c}^{-1}]^{I} ;  \, \,\,\,\,\,\,\,\,
           \widetilde{\Gamma}_{c}^{I}= [ \widetilde{\Pi}_{0c}^{-1} ]^{I} + \alpha [ \widetilde{\Pi}_{c}^{-1} ]^{I}
             .     \nonumber \\
  &&   \label{55}
\end{eqnarray}

Equation (55) holds for each cluster index $I$, and the $\widetilde{G}_{c}^{I}$ and $\widetilde{\Pi}_{c}^{I}$
 are determined by
$ [ \widetilde{G}_{0c}^{-1} ] ^{I}$ and $ [ \widetilde{\Pi}_{0c}^{-1} ]^{I}$ of the reference system
via Eqs. (40)$-$(45). Since the matrices on both sides of Eq. (55) are block diagonal, the cluster
index $I$ can be dropped.

This set of equations apply in the case where the reference system has
continuous degrees of freedom, such as a cluster where each site is coupled to
continuous baths. It can be regarded as an extensions of the EDMFT toward the cluster algorithm.\cite{Sun1}
Note that in this formula, the clusters of the reference system have open boundary
conditions like CDMFT. We therefore denote this extension as EDMFT$+$CDMFT.
It will generally lead to self-energies that are not spatially translation invariant. As in
CDMFT, the self-energies for the lattice model should be reevaluated from the cluster self-energies
obtained from Eq. (55). Different schemes of estimators have been discussed in Ref. 13.
When the solutions are confined to be spatial translation invariant, each cluster becomes identical.
In particular, for the reference system composed of one impurity
embedded in two baths described by local Weiss field $\widetilde{G}_{0}^{-1}$ and $\widetilde{\Pi}_{0}^{-1}$,
the original EDMFT equations will be recovered. In this case, the right hand of the first two lines of
Eq. (55) can be evaluated in momentum space, and the local Green's functions
are written as integrals with free density of states. The EVCA equations reduce to

\begin{eqnarray}
\widetilde{G} \left( i \omega_n \right) &=& \int \frac{\rho_{0} (\epsilon) }{ i \omega_n + \mu -\epsilon
        - \widetilde{\Sigma}\left( i \omega_{n} \right) } d \epsilon   , \nonumber \\
\widetilde{\Pi} \left( i \nu_{n} \right) &=& -\alpha \int \frac{V_{0} (\epsilon) }{ - \epsilon
        - \widetilde{\Gamma}\left( i \nu_{n} \right) } d \epsilon       ,  \nonumber \\
 \widetilde{\Sigma}\left( i \omega_n \right) &=&  \widetilde{G}_{0}^{-1}\left( i \omega_n \right)
     - \widetilde{G}^{-1}\left( i \omega_n \right)  ,  \nonumber \\
  \widetilde{\Gamma}_{c} \left( i \nu_{n} \right) &=&
     \widetilde{\Pi}_{0}^{-1}\left( i \nu_{n} \right)  + \alpha  \widetilde{\Pi}^{-1} \left( i \nu_{n} \right)
	. \label{56}
 \end{eqnarray}
Quantities here are the local ones, and
$\rho_{0} (\epsilon)= 1/N \sum_{k} \delta (\epsilon - \epsilon_k)$,
$V_{0} (\epsilon)= 1/N \sum_{k} \delta (\epsilon - V_k)$ are the density distributions of the
eigenvalues of hopping matrix and density-density interaction matrix. $\omega_n$
and $\nu_n$ are the fermionic and bosonic Matsubara frequencies, respectively. $\epsilon_k$
and $V_k$ are obtained from Fourier transformation of the hopping and Coulomb interaction parameters
in the original Hamiltonian,
\begin{eqnarray}
   \epsilon_{k} &=& \sum_{i-j} -t_{ij} e^{-i {\bf k} \left({\bf r}_i - {\bf r}_j \right) }  , \nonumber \\
   V_{k} &=& \sum_{i-j} V_{ij} e^{-i {\bf k}  \left({\bf r}_i - {\bf r}_j \right) }      . \label{57}
\end{eqnarray}

If the reference system has only finite degrees of freedom,
the variation of $\{ \widetilde{\Sigma}_{c}, \widetilde{\Gamma}_{c} \}$ cannot
be arbitrary. We can use a finite number of variational parameters $ \widetilde{t}$ and $\widetilde{V}$
to specify the Green's functions $\widetilde{G}_{0c}^{-1} ( \widetilde{t})$ and
$\widetilde{\Pi}_{0c}^{-1} ( \widetilde{V} )$, respectively.
The stationary point of $\widetilde{\Omega}_{EVCA}
[ \widetilde{\Sigma}_{c}, \widetilde{\Gamma}_{c} ]$ may be
unreachable in the space of $\{ \widetilde{t}, \widetilde{V} \}$, which corresponds
to a certain subspace of  $\{ \widetilde{\Sigma}_{c}, \widetilde{\Gamma}_{c} \} $.
 In such cases, we identify the approximate self-energies
 of the original system as the ones at the stationary point of the functional
$\widetilde{\Omega}_{EVCA}
[ \widetilde{\Sigma}_{c} ( \widetilde{t}, \widetilde{V}),
\widetilde{\Gamma}_{c}( \widetilde{t}, \widetilde{V} )]$,

\begin{eqnarray}
  && \frac{\delta \widetilde{\Omega}_{EVCA}
  \left[ \widetilde{\Sigma}_{c}\left( \widetilde{t}, \widetilde{V} \right) ,
  \widetilde{\Gamma}_{c} \left( \widetilde{t}, \widetilde{V} \right) \right]}
  {\delta \widetilde{t} }
  |_{\widetilde{t}=t_{app}, \widetilde{V}=V_{app}}  =0 ,    \nonumber \\
&& \frac{\delta \widetilde{\Omega}_{EVCA}
 \left[\widetilde{\Sigma}_{c}\left( \widetilde{t}, \widetilde{V} \right) ,
 \widetilde{\Gamma}_{c}\left( \widetilde{t}, \widetilde{V} \right)  \right]}
{\delta \widetilde{V}}
|_{ \widetilde{t} =t_{app}, \widetilde{V}=V_{app}}  =0             .  \nonumber   \\
&&                \label{58}
\end{eqnarray}
It further simplifies into
\begin{eqnarray}
  &&   {\text Tr} \left[ \widetilde{G}_{c} - \left|  \left( G_{0}^{-1} - \widetilde{\Sigma}_{c}  \right)^{-1} \right| \right]
            \frac{\delta \widetilde{\Sigma}_{c}\left( \widetilde{t}, \widetilde{V} \right) }{\delta \widetilde{t}}  \nonumber \\
  &&  +{\text Tr} \left[ \widetilde{\Pi}_{c} + \alpha \left|  \left( \Pi_{0}^{-1} - \widetilde{\Gamma}_{c}  \right)^{-1} \right| \right]
             \frac{\delta \widetilde{\Gamma}_{c}\left( \widetilde{t}, \widetilde{V} \right) }{\delta \widetilde{t}}
	     =0        ,  \nonumber \\
  &&	        \label{59}
\end{eqnarray}
\begin{eqnarray}
 &&   {\text Tr} \left[ \widetilde{G}_{c} - \left|  \left( G_{0}^{-1} - \widetilde{\Sigma}_{c}  \right)^{-1} \right| \right]
            \frac{\delta \widetilde{\Sigma}_{c}\left( \widetilde{t}, \widetilde{V} \right) }{\delta \widetilde{V}}  \nonumber \\
  &&  +{\text Tr} \left[ \widetilde{\Pi}_{c} + \alpha \left|  \left( \Pi_{0}^{-1} - \widetilde{\Gamma}_{c}  \right)^{-1} \right| \right]
             \frac{\delta \widetilde{\Gamma}_{c}\left( \widetilde{t}, \widetilde{V} \right) }{\delta \widetilde{V}}
	     =0        .   \nonumber  \\
  &&           \label{60}
\end{eqnarray}

Equation (58) is only one of the many possible ways to construct the discrete version of
the EVCA equations. In Eqs. (59)$-$(60), the fermion and boson contributions are combined
through the cross derivation terms
$\delta \widetilde{\Gamma}_{c}\left( \widetilde{t}, \widetilde{V} \right) / \delta \widetilde{t} $
and $\delta \widetilde{\Sigma}_{c}\left( \widetilde{t}, \widetilde{V} \right) / \delta \widetilde{V}$.
We can also conceive another form of discrete EVCA which has the same
continuous limit as Eq. (55), but without the cross derivative terms.
For example, we propose the following discrete version of EVCA which
 separates the contributions from fermions and bosons.
Define
\begin{eqnarray}
  && \widetilde{\Omega}_{f}\left[\widetilde{\Sigma}_{c}, \widetilde{\Gamma}_{c}\right]
  = \widetilde{\Omega} \left[\widetilde{\Sigma}_{c}, \widetilde{\Gamma}_{c} \right]  \nonumber \\
 && +  \frac{1}{\beta} {\text Tr} \ln \left(\widetilde{G}_{0c}^{-1} -\widetilde{\Sigma}_{c}\right)
    - \frac{1}{\beta} {\text Tr} \ln \left(G_{0}^{-1} -\widetilde{\Sigma}_{c}\right)   ;  \nonumber \\
  && \widetilde{\Omega}_{b}\left[\widetilde{\Sigma}_{c}, \widetilde{\Gamma}_{c}\right]
  = \widetilde{\Omega} \left[\widetilde{\Sigma}_{c}, \widetilde{\Gamma}_{c} \right]  \nonumber \\
  && - \frac{\alpha}{\beta}{\text Tr} \ln \left(\widetilde{\Pi}_{0c}^{-1}- \widetilde{\Gamma}_{c} \right)
        + \frac{\alpha}{\beta} {\text Tr} \ln \left(\Pi_{0}^{-1}- \widetilde{\Gamma}_{c} \right)   ,       \label{61}
\end{eqnarray}
we get a new version of the EVCA equations
\begin{eqnarray}
  && \frac{\delta \widetilde{\Omega}_{f}
  \left[ \widetilde{\Sigma}_{c}\left( \widetilde{t}, \widetilde{V} \right) ,
  \widetilde{\Gamma}_{c} \left( \widetilde{t}, \widetilde{V} \right) \right]}
  {\delta \widetilde{t} }
  |_{\widetilde{t}=t_{app}}  =0  ,   \nonumber \\
&& \frac{\delta \widetilde{\Omega}_{b}
 \left[\widetilde{\Sigma}_{c}\left( \widetilde{t}, \widetilde{V} \right) ,
 \widetilde{\Gamma}_{c}\left( \widetilde{t}, \widetilde{V} \right)  \right]}
{\delta \widetilde{V}}
|_{\widetilde{V}=V_{app}}  =0              .       \label{62}
\end{eqnarray}
This approximation scheme leads to the equations
\begin{eqnarray}
  &&   {\text Tr} \left[ \widetilde{G}_{c} - \left|  \left( G_{0}^{-1} - \widetilde{\Sigma}_{c}  \right)^{-1} \right| \right]
            \frac{\delta \widetilde{\Sigma}_{c}\left( \widetilde{t}, \widetilde{V} \right) }{\delta \widetilde{t}}  =0 \nonumber \\
  &&  {\text Tr} \left[ \widetilde{\Pi}_{c} + \alpha \left|  \left( \Pi_{0}^{-1} - \widetilde{\Gamma}_{c}  \right)^{-1} \right| \right]
             \frac{\delta \widetilde{\Gamma}_{c}\left( \widetilde{t}, \widetilde{V} \right) }{\delta \widetilde{V}}
	     =0        .    \nonumber \\
  &&   \label{63}
\end{eqnarray}

Both Eqs. (59) and (60) and Eq. (63) are extensions of the VCA \cite{Potthoff1} to include the nonlocal
 density-density interactions beyond Hartree approximation. They recover the continuous version Eq. (55)
 in the continuous bath limit. It is noted that Eqs. (59) and (60) are formally exact, while Eq. (63) is not.
 Equation (63) can be
 regarded as a further approximation based on Eqs. (59) and (60) to neglect the cross-derivative terms.
 In our numerical study of the three-site EVCA, however, it is found that due to the very large energy
  scale difference between fermion and boson contributions to the grand potential, in the first version of EVCA,
  the cross-derivative terms will spoil the self-consistent equations and make it difficult to find the stationary point.
  The second version, Eqs. (61) $-$ (63), works much better. As shown in the next section,
  it gives out rather satisfactory results for the extended Hubbard model.

In the above theories, it is rather flexible to select the reference system according to
the nature of the problem at hand. One can select the size and the shape of the cluster,\cite{Dahnken1,
Aichhorn1} the number
of discrete bath sites, as well as the definition of the variational parameters used in the reference system.
For comparable complexity of the reference system, it is expected that the lower spatial dimensions
the system has, the more efficient it is to use larger size of the cluster instead of to use larger
number of bath sites. In the VCA study of the Mott-Hubbard transition,\cite{Potthoff1,Pozgajcic1}
qualitatively correct results have already been obtained by using a reference system
 in which each decoupled cluster contains only two sites, one impurity site and one fermion bath site.
This shows the amazing efficiency of the VCA. It is expected that EVCA may share
these advantages of the VCA, and becomes a powerful technique for the study of strongly correlated
electron systems. Interesting physical problems, such as different phases in the extended
Hubbard model, the competition between the Kondo and Ruderman-Kittel-Kasuya-Yosida (RKKY) interaction, etc.,
may be studied by EVCA using small reference systems.

\subsection{Three-site EVCA: An example}

In this subsection, as an example, we present some numerical results for the simplest realization of the EVCA for
the extended Hubbard model.
Chitra {\it et al.} studied the influence of the nonlocal Coulomb repulsion on the Mott metal-insulator transition
using EDMFT and the projection technique.\cite{Chitra2}
They found that the long-range Coulomb repulsion will change the second-order Mott transition
at zero temperature into a first-order transition.
Here we use the second version of the discrete EVCA, Eqs. (62) and (63), to illustrate the implementation
of EVCA, and focus only on the properties of the metal and the insulator.
For simplicity, we only consider the half-filling and paramagnetic phase.
More detailed studies on the extended Hubbard model will be published elsewhere.
The Hamiltonian that we will study is Eq. (3),
\begin{eqnarray}
H & = & -\sum\limits_{i,j, \sigma } t_{i j}c_{i\sigma }^{\dagger}c_{j\sigma }
             + \sum\limits_{ i,j }V_{i j} n_{i}n_{j}-\mu \sum_{i}n_{i} \nonumber \\
     & &  + U\sum_{i} n_{i \uparrow} n_{i \downarrow}   \,\, .          \label{64}
\end{eqnarray}
$G_{0}^{-1}$ and $\Pi_{0}^{-1}$ are therefore the same as in Eqs. (5) and (6).
We consider a reference system composed of disconnected lattice sites, each of which
is coupled with two bath sites: one fermionic site and one bosonic site. The unit cell of the reference system is
schematically shown in Fig. 1. This is an extension of the two-site DMFT that works well on a qualitative level
in the study of the Mott-Hubbard transition.\cite{Potthoff1}

\begin{figure}[ht]
%\vspace{-0.3cm}
\includegraphics[width=3.0cm, height=2.5cm]{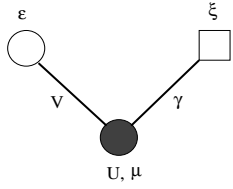}
\begin{center}
%\psfig{figure=3site.eps,width=0.15\textwidth}
%\vspace{0.3cm}
\caption{\label{fig:site}Structure of a three-site unit cell of the reference system Hamiltonian (66).
Filled dot denotes impurity site, open dot and square denote fermion and boson bath site,
respectively.}
\label{fig1}
\end{center}
\end{figure}

%\begin{figure}
%\includegraphics{fig_1}% Here is how to import EPS art
%\caption{\label{fig:epsart} A figure caption. The figure captions are
%automatically numbered.}
%\end{figure}

In the following part of this section, we will drop the tilde on the top of the variables $\widetilde{G}$,
$\widetilde{\Pi}$, etc., since this will not introduce confusion.
For the reference system described above, the action Eq. (41) on a single site is written as
\begin{eqnarray}
S_{ref} &=& \sum_{\sigma} \int_{0}^{\beta} d \tau \int_{0}^{\beta} d\tau^{\prime} c_{\sigma}^{*} \left(\tau \right)
                    \left[ -G_{0\sigma}^{-1}\left( \tau- \tau^{\prime} \right)\right] c_{\sigma} \left(\tau^{\prime} \right)
		     \nonumber \\
	   &+&  \int_{0}^{\beta} d \tau \int_{0}^{\beta} d\tau^{\prime}
		     n(\tau) \left[- \Pi_{0}^{-1}\left( \tau- \tau^{\prime} \right)\right] n(\tau^{\prime}) \nonumber \\
	   &+& U \int_{0}^{\beta}d\tau n_{\uparrow}(\tau) n_{\downarrow}(\tau)
	            + 2p\int_{0}^{\beta}d\tau n(\tau) - \beta p \langle n \rangle      .     \nonumber \\
  &&      \label{65}
\end{eqnarray}
Here $p=[ 1/\Pi_{0} \left( i \nu_{0}\right)+ \sum_{j} V_{0j} ] \langle n \rangle$.
The Hamiltonian for the three-site unit cell of the reference system (Fig. 1) can be written as
\begin{eqnarray}
  H_{imp} &=& \epsilon \sum_{\sigma} a_{\sigma}^{\dagger} a_{\sigma}
                       + V \sum_{\sigma} \left(a_{\sigma}^{\dagger} c_{\sigma}
                       +  c_{\sigma}^{\dagger} a_{\sigma} \right) + Un_{\uparrow}n_{\downarrow}    \nonumber \\
		 &-& (\mu-2p)\sum_{\sigma}c_{\sigma}^{\dagger} c_{\sigma}
		       + \xi b^{\dagger}b+ \gamma \left( b^{\dagger}+b\right) \sum_{\sigma}c_{\sigma}^{\dagger} c_{\sigma} .
		           \nonumber \\
  &&         \label{66}
\end{eqnarray}
Here $n_{\sigma}=c_{\sigma}^{\dagger}c_{\sigma}$. $c_{\sigma}$ and
$a_{\sigma}$ are the annihilation operators on impurity and fermion bath sites, respectively.
$b$ is the boson annihilation operator on the boson bath site.
Integrating out the bath degrees of freedom in $H_{imp}$, and comparing the resulting effective impurity action with
Eq. (65), we obtain the following relations:
\begin{eqnarray}
    && G_{0 \sigma}^{-1} \left( i \omega_{n}\right)= i \omega_{n}+ \mu - \frac{V^2}{i\omega_{n}-\epsilon} ,
                \nonumber \\
    && \Pi_{0}^{-1}\left( i \nu_{n}\right)= \frac{\gamma^{2} \xi}{\nu_{n}^{2}+ \xi^{2}}      ,    \label{67}
\end{eqnarray}
\begin{eqnarray}
    \Omega_{ref} &=& -\frac{1}{\beta} \ln \prod_{\sigma} \int {\mathcal D} c_{\sigma}^{*}(\tau) \int {\mathcal D}
                                                  c_{\sigma}(\tau) e^{-S_{ref}}  \nonumber \\
                                             &=& -\frac{1}{\beta} \ln \left[ {\text Tr} e^{-\beta H_{imp}} \right] +\frac{2}{\beta}
					             \ln \left(1+e^{-\beta\epsilon} \right)  \nonumber \\
				          && - \frac{1}{\beta} \ln \left(1-e^{-\beta\xi} \right)-p \langle n\rangle    .   \label{68}
\end{eqnarray}

The expression for the functional $\Omega_{f}$ and $\Omega_{b}$ in Eq. (61) now reduces to\cite{note4}
\begin{eqnarray}
 && \frac{1}{N} \Omega_{f}    \nonumber \\
 &=& \Omega_{ref} - \frac{4}{\beta} \sum_{n=0}^{+\infty}
                 \int d\epsilon \, \rho_{0} (\epsilon) \ln \left|1+ G \left(i \omega_{n} \right)
		 [\Delta(i \omega_n)- \epsilon ] \right| ,    \nonumber \\
&& \frac{1}{N} \Omega_{b}   \nonumber \\
&=&  \Omega_{ref} +\frac{2 \alpha}{\beta}\sum_{n=1}^{+\infty}
                 \int d\epsilon \, V_{0} (\epsilon) \ln \left |1- \frac{1}{\alpha} \Pi \left(i \nu_{n} \right)
		 [\Phi(i \nu_n)- \epsilon ] \right|     \nonumber \\
                &+& \frac{\alpha}{\beta}
                 \int d\epsilon \, V_{0} (\epsilon) \ln \left|1- \frac{1}{\alpha} \Pi \left(i \nu_{0} \right)
		 [\Phi(i \nu_0)- \epsilon ] \right|     , \label{69}
\end{eqnarray}
where
\begin{eqnarray}
    && \Delta(i \omega_n)=  \frac{V^2}{i \omega_n - \epsilon} ,  \nonumber \\
    &&  \Phi(i \nu_n) = \frac{\gamma^2 \xi}{\nu_n^{2} + \xi^{2}}  .            \label{70}
\end{eqnarray}

The Green's functions $G$ and $\Pi$ can be obtained through the Lehmann expression
after the three-sites Hamiltonian Eq. (66) is diagonalized numerically. They are
\begin{equation}
  G_{\sigma} \left( i \omega_l \right) = \frac{1}{\Xi_{imp}} \sum_{mn}
                  \frac{|\langle n |c_{\sigma}^{\dagger}|m \rangle|^2 }{i \omega_l +E_m - E_n }
		  \left( e^{-\beta E_m} + e^{- \beta E_n}\right)         \label{71}
\end{equation}
and
\begin{eqnarray}
  && \Pi \left( i \nu_l \right)  \, (l \ne 0)   \nonumber \\
  &=&  \frac{1}{\Xi_{imp}} \sum_{mn}
                  \frac{|\langle n |n| m \rangle|^2 }{i \nu_l +E_n - E_m }
		  \left( e^{-\beta E_m} - e^{- \beta E_n}\right)     ,   \nonumber  \\
  &&		  \label{72}
\end{eqnarray}
\begin{eqnarray}
   && \Pi \left( i \nu_0 \right)  \nonumber \\
   &=& \frac{\beta}{\Xi_{imp}} \sum_{mn, \left( E_m=E_n \right)}
                 |\langle n |n| m \rangle|^2 e^{-\beta E_m}           \nonumber \\
   &+& \frac{1}{\Xi_{imp}} \sum_{mn, \left( E_m \ne E_n \right) }
		  \frac{|\langle n |n| m \rangle|^2 }{E_n - E_m }
		  \left( e^{-\beta E_m} - e^{- \beta E_n}\right)    .  \nonumber \\
  &&		    \label{73}
\end{eqnarray}
Here $\Xi_{imp}= {\text Tr} e^{-\beta H_{imp}}$.

The densities of states of the matrices hopping $t_{ij}$ and interaction $V_{ij}$ are selected to be
 of semicircular form,
\begin{eqnarray}
     &&  \rho_{0} (\epsilon) = \frac{2}{\pi W^2} \sqrt{W^2- \epsilon^2}  ,   \nonumber \\
     &&  V_{0} (\epsilon) = \frac{2}{\pi V_{0}^2} \sqrt{V_{0}^2- \epsilon^2}   .   \label{74}
\end{eqnarray}
These densities of states can be realized by nearest-neighbor couplings $t_{ij}$ and $V_{ij}$
on the Bethe lattice in the limit of the coordination number $z \rightarrow \infty$, with scaling
$t \rightarrow W/2\sqrt{z}$ and $V \rightarrow V_0/2\sqrt{z}$, respectively.
It is noted that within the present formula, the description of the nonlocal Coulomb repulsion $V_{ij}$
is only through two quantities $V_{0}(\epsilon)=1/N \sum_{k} \delta \left( V_{k} - \epsilon \right)$
and $\sum_{j}V_{0j}=V_{k=0}$. It is apparent that the description
is far from complete. A complete description should take into account different contributions from
each $k$ component of $V_{ij}$, which can be realized partly by using larger a cluster in the reference system.

Another point is that using the above scaling and in the limit $z \rightarrow \infty$,
we get a diverging $V_{k=0}$. This reflects the fact that our present EVCA formula for density-density interaction
is not a well-defined theory for infinite-dimensional models. The same problem exists for the symmetry-broken EDMFT.
However, this doesnot prevent the theory from being a good approximation for finite-dimensional systems.
In our calculation, we simply take $V_{k=0}$ as an independent parameter for the original system. This parameter
can be absorbed into the definition of $\mu$, and does not play a role in the half-filled case. We simply set
$V_{k=0}=1$ in the calculation.

The numerical aspect of implementing the VCA has been studied in detail.\cite{Potthoff1,Pozgajcic1}
Here we use a Matsubara frequency formalism, and the calculation is composed of several steps.
First, diagonalize the three-site Hamiltonian
Eq. (66), and calculate the Green's functions $G$
and $\Pi$ through Eqs. (71)$-$(73). Second, $1/N\Omega_{f}$ and $1/N\Omega_{b}$
in Eq. (69) are evaluated. Finally, we search for the point in the space of $\{ \epsilon, V, \xi, \gamma \}$ where
$1/N\Omega_{f}$ is stationary with respect to the fermion parameters
$\epsilon$, $V$, and $1/N\Omega_{b}$ with respect to the boson parameters $\xi, \gamma$.
The Green's functions at such stationary points are our solutions for the three-site EVCA.
For the boson operator $b$, we take $N_b < \infty$ optimized orthogonal
 states as the bases.\cite{Bulla2} The final results should converge with respect to $N_{b}$.
 For the half-filled system that we consider here, the particle-hole symmetry condition of the reference
Hamiltonian Eq. (66) is
\begin{eqnarray}
 && \mu=\frac{U}{2}- \frac{2 \gamma^2}{\xi} +2 p   \nonumber \\
 &&  \epsilon=0  \,\, \, \,\, \, \, \,  N_b= \infty    .  \label{75}
\end{eqnarray}
It is found that with $N_{b}=8$, the particle-hole symmetry is already
satisfied very well, and the results do not change on further increasing $N_b$.
With this $N_b$ and considering the conservative quantity
$n_{\uparrow}$ and $n_{\downarrow}$, the size of the largest matrix to be diagonalized is 48. This enable us
to scan the parameter space quite quickly. For the parameter $\alpha$  that is unfixed within our theory,
we take the value $\alpha=1/2$, same as in other EDMFT studies.\cite{note3}
The influence of different selection of $\alpha$ will be discussed elsewhere.

\begin{figure}[ht]
\vspace{0.7cm}
\begin{center}
\includegraphics[width=5.5cm,height=6.5cm]{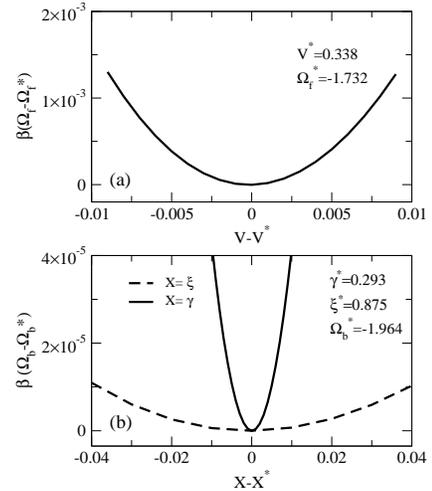}
\vspace{0.3cm}
\caption{The $\Omega_{f}$ and $\Omega_{b}$ in the vicinity of the metallic stationary point at $T=0.01$, $U=1.0$,
and $V_{0}=0.5$. They are shifted by their stationary values $\Omega_{f}^{*}$ and $\Omega_{b}^{*}$, respectively, and
magnified by $\beta=100$. $V^{*}$, $\xi^{*}$, and $\gamma^{*}$ denote the corresponding values at the
stationary point.}
\label{fig2}
\end{center}
\end{figure}

In the following, we summarize our results for the extended Hubbard model. Since the main purpose is to
check the theory, the study is limited to half filling and in the paramagnetic, charge-disordered phase.
At low temperatures, a critical $U_c$ value separates the metallic and the Mott insulating phases.
The nonlocal Coulomb repulsion $V_{0}$ may influence the $U_c$ value, and may even change the nature
of the phase transition. For $V_{0}$ sufficiently large, a charge-ordering phase becomes more stable. The
description of such ordered phase needs the symmetry-broken version of EVCA.
Our discussion below resides in the metallic and insulating phase separately, and leave the detailed study of the
Mott transition and charge-ordering transition for later work.

We set $W=1.0$ as
the energy unit. All the data shown here are obtained for a temperature $T=0.01 < T_c$,
where $T_c \approx 0.014$ is the critical temperature of the first-order Mott transition obtained for the two-site VCA.
In Fig.2, an example is shown for the stationary point of $\Omega_{f}$ and $\Omega_{b}$ at U=1.0
which is on the metal side of the Mott transition.
Since we fixed $\epsilon=0$, there are three free variational parameters left,
$\gamma$, $V$, and $\xi$.
It is a metallic solution with finite $V^{*}$ values. The vertical scales in Fig. 2(a) and 2(b) shows
that $\Omega_{f}$ changes two orders of magnitude larger
 than $\Omega_{b}$, given similar variation of the parameters. This is a hint of why the first version of
 EVCA (58) works not so well in this case and we have to adopt the EVCA Eq. (62).
Now we check to what extent the above three-site EVCA solution is close to the EDMFT solution.
In the continuous bath limit, the EVCA Eq. (56) with semicircular density of states (74)
will lead to the following EDMFT relations:
\begin{eqnarray}
\Delta\left( i \omega_n \right) =\frac{W^2}{4} G_{\sigma}\left( i \omega_n \right)
= \frac{W^2}{4} X_{\sigma}\left( i \omega_n \right)   , \nonumber \\
\Phi\left( i \nu_n \right) =\frac{V_{0}^2}{4 \alpha} \Pi\left( i \nu_n \right)
= -\frac{V_{0}^2}{4} Y\left( i \nu_n \right)   ,  \label{76}
\end{eqnarray}
where
\begin{eqnarray}
   && X_{\sigma}\left( i \omega_n \right)=\int \frac{\rho_{0}\left(\epsilon \right)}{\Delta\left( i\omega_n\right) - \epsilon
                   + G_{\sigma}^{-1}\left( i\omega_n\right)}  d\epsilon  ,  \nonumber \\
  && Y\left( i \nu_n \right)=\int \frac{V_{0}\left(\epsilon \right)}{-\alpha \Pi^{-1} \left( i \nu_n \right)-\Phi \left( i \nu_n \right)
                                       -\epsilon}  d\epsilon    .  \label{77}
\end{eqnarray}
When the baths have only finite degrees of freedom, the solutions of EVCA Eq. (63) will be different from
the continuous limit solutions, and hence Eq. (76) will not be satisfied anymore.
Therefore the discrepancies between the three quantities can be taken as a criterion
for the quality of the three-site EVCA solution.
In Fig. 3(a), the solution of the Green's functions $\Pi \left(i\nu_n \right)$ at the stationary point for $T=0.01, U=1.0$ is shown.
The differences between $ (V_{0}^2 / 4\alpha) \Pi\left( i \nu_n \right)$ and $\Phi \left( i \nu_n \right)$,
and that between $(V_{0}^2/4\alpha) \Pi\left( i \nu_n \right)$ and $-(V_{0}^2/4 )Y\left( i \nu_n \right)$ are
shown in Fig. 3(b).
\begin{figure}[ht]
\vspace{0.4cm}
\begin{center}
\includegraphics[width=5.7cm,height=5.5cm]{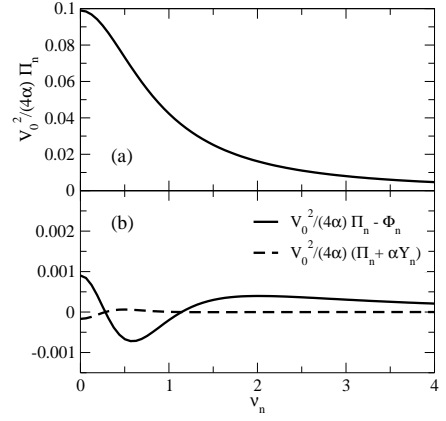}
\vspace{0.3cm}
\caption{(a) The density-density Green's function at the stationary point for $T=0.01$, $U=1.0$, and
$V_{0}=0.5$; (b) the discrepancies between certain quantities, which should be zero
in the continuous limit of bath, see text. $\alpha=1/2$.}
\label{fig3}
\end{center}
\end{figure}
It is seen that the fulfillment of Eq. (76) is rather good for the present case, although we have only three
variational parameters. The relative discrepancy in the small-frequency regime is less than $1\%$.
This shows that the EVCA indeed has the similar merit of the VCA, i.e., it is very efficient to represent the
continuous baths degrees of freedom by a small cluster. For other $V_0$ values, we observed that
with $V_{0}$ increasing, the relative discrepancies also increases. Up to the upper boundary $V_{max}$ of the
parameter $V_0$, the relative discrepancies are less than $10\%$.

For fixed $U$ and $T$, the upper boundary $V_{max}$ is defined as the value of $V_0$ above which
 the pole in the integral expression of $Y\left(i \nu_n \right)$ [second equation of Eqs. (77)] enters the
  range $[-V_0, V_0 ]$. This will lead to a singular peak in $Y \left( i \nu_n \right)$ and the EDMFT
  equation (76) no longer has a solution. This problem of EDMFT was observed previously.\cite{Sun1,Smith1}
  In the EVCA formula, it appears in the second expression of Eq. (69) where the absolute value is used in
  the argument of logarithm. This does not change the stationary point, but can formally extend the formalism
   to all regime of $V_0$. In the regime $V_0 > V_{max}$ however, we find no physical EVCA stationary point,
   which is consistent with the nonexistence of an EDMFT solution in this regime. This breakdown of the present
   formalism in the large-$V_0$ regime is related to the instability of the symmetry-unbroken solutions.
It is expected that in the regime $V_0 > V_{max}$, a symmetry-broken solution should be stable
in the corresponding EVCA equations.\cite{Sun1}

\begin{figure}[ht]
\vspace{0.5cm}
\begin{center}
\includegraphics[width=4.9cm,height=3.8cm]{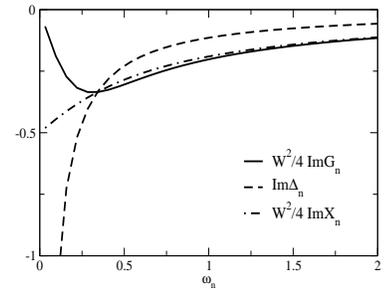}
\vspace{0.3cm}
\caption{Comparison of the three quantities related to the electron Green's function
at the stationary point for $T=0.01$, $U=1.0$, and $V_{0}=0.5$. }
\label{fig4}
\end{center}
\end{figure}

In Fig. 4, we compare the three quantities for the electrons,
 $W^2/4 {\text Im}G_{\sigma}\left(i \omega_n \right)$, ${\text Im}\Delta\left(i \omega_n \right)$,
 and $W^2/4{\text Im} X_{\sigma}\left(i \omega_n \right)$ at the same stationary point as in Fig. 3.
 Since there is only one pole $\omega_n=0$ in the hybridization function $\Delta\left(i \omega_n \right)$,
 the discrepancies among them enlarge in the small-frequency limit. However, they are fairly small
  when the frequency is larger than the common crossing point.

In both Fig. 3 and 4, it is interesting to observe that there are certain common crossing
points among the three quantities. In each subequation of Eq. (76), the two successive
 identities are not independent. One implies the other. Therefore the three quantities compared in Fig. 3(b)
 as well as in Fig. 4 always cross at common points. There are two frequency values of
 $\nu$ and one of $\omega$, at which the first and second equation of Eq. (76) are satisfied,
 respectively. This corresponds to the fact that we have two free parameters $\xi, \gamma$ for bosons
 and one free parameter $V$ for fermions in the three-site unit cell at half filling. Therefore,
 the EVCA algorithm can also be regarded as a specific way of fixing a number of
boson and fermion Matsubara frequencies on which the EDMFT equation are satisfied exactly.
 In the limit of infinite degrees of freedom of the reference system, the EDMFT equations
 are then fulfilled at all frequencies. This again shows that EDMFT is the continuous limit of one-impurity EVCA.
 This observation also provides a possibility of constructing theories parallel to the EVCA. One can
  conceive certain ways of selecting the Matsubara frequencies at which the EDMFT equation is satisfied exactly.
  The number of such frequencies represents the degrees of freedom of the reference system and can be increased
   in a controlled way. However, a unique criterion for the qualities of different such approaches is still lacking.

\begin{figure}[ht]
\vspace{1.0cm}
\begin{center}
\includegraphics[width=6.5cm,height=4.7cm]{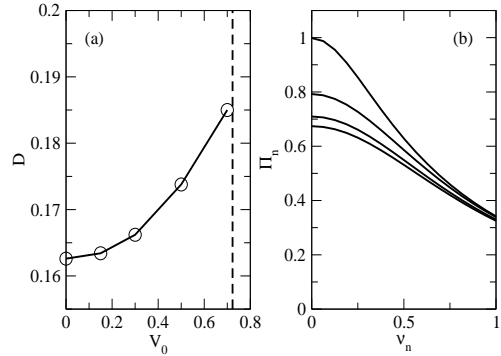}
\vspace{0.3cm}
\caption{(a) Double occupancy as a function of $V_0$ for $T=0.01$, $U=1.0$.
The vertical dashed line marks out the boundary $V_{max} \approx 0.72$,
above which the symmetry-unbroken three-site EVCA solution doesnot exist.
(b) The density-density Green's function $\Pi\left(i \nu_n \right)$ for several $V_0$ values.
From top to bottom, $V_0=0.7, 0.5, 0.3$, and $0.0$.}
\label{fig5}
\end{center}
\end{figure}

For $T=0.01$ and $U=1.0$, the metallic symmetry-unbroken solution of EVCA persists up
to $V_{max} \approx 0.72$. With increasing interaction strength $V_0$, the density-density Green's function
$\Pi\left( i\nu_n \right)$ increases uniformly on all energy regime, as illustrated in Fig. 5(b).
This shows that the charge fluctuations in the metallic phase are enhanced by the nonlocal interactions.
At half filling, the total weight of $\Pi\left( i\nu_n \right)$ is given by
$\beta \left(\langle n \rangle - \langle n \rangle^2 +2 D\right)= 2 \beta D$,
where $D=\langle n_{\uparrow} n_{\downarrow}\rangle$
 is the double occupancy.
 In Fig. 5(a), we show the double occupancy $D$ as a function of $V_0$ for $U=1.0$ and $T=0.01$.
 It is an increasing function up to $V_{max}$ where the symmetry-unbroken solution disappears.
 Physically, the enhancement of the charge fluctuations is related to the dynamical screening of the local
 repulsion by the nonlocal interactions. This effect has been observed in previous EDMFT
 studies of the extended Hubbard model,\cite{Sun1} and may have qualitative influence on the
  metal-insulator phase transition.\cite{Chitra2}

In the insulating phase, the charge fluctuations are suppressed to very much extent,
and the effect of nonlocal interaction is almost negligible at low temperatures.
Our calculation at $T=0.01, U=3.2$ gives a stationary point $V^{*}=0.04$, $\xi^{*}=1.1$,
 and $\gamma^{*}=0.012$. The double occupancy at this stationary point is $D=3.6 \times 10^{-4}$. It is expected
 that the finite $V^{*}$ here is due to the finite temperature that we use. At exactly zero temperature,
  the two-site VCA gives zero double occupancy and vanishing $\Pi\left(i \nu_n \right)$.
  This will lead to vanishing of the Weiss field $\Pi^{-1}_{0}$ according to the EVCA equations
  Eq. (76). Therefore, we expect that for the three-site EVCA, the finite $V_0$ does not play a role
  in the zero-temperature insulating phase, and the results are the same as in the pure
  Hubbard model. Of course, this is an artifact of the three-site EVCA. It is known that for the Hubbard
   model, the double occupancy $D$ is finite even in the insulating solution at zero temperature.
   More detailed study on the effect of nonlocal interaction on the metallic and insulating phases and
    Mott metal-insulator transition is to be carried out systematically in the EVCA approach.

\section{Correlated Hopping}
In this section, we construct EVCA for the Hubbard model with correlated hopping.
The correlated hopping term is expected to play an important role in the phenomenon of
itinerant ferromagnetism. \cite{Kollar1}
We consider the following Hamiltonian, \cite{Shvaika1}

\begin{equation}
  H=- \sum_{ij \sigma} \sum_{m,n=1}^{2} t_{ij}^{mn}
  c_{i \sigma}^{\dagger} P_{i \sigma}^{m} c_{j \sigma} P_{j \sigma}^{n} + H_{loc}  , \label{78}
\end{equation}
where $P_{i \sigma}^{1}=n_{i \overline{\sigma}}$ and $P_{i \sigma}^{2}=1- n_{i \overline{\sigma}}$, and
 $H_{loc}= U \sum_{i} n_{i \uparrow} n_{i \downarrow} $ is the local Hubbard interaction. In Eq. (78),
  the most general form
of correlated hopping is described by the hopping parameter $t_{ij}^{mn}$.
For this model, the DMFTCH has already been developed,\cite{Schiller2,Shvaika1}
 where the lattice model is
mapped into one impurity embedded into several baths. Each bath is coupled to a different
component of the impurity degrees of freedom. The EVCA formula to be established in the following
is an extension of DMFTCH to larger cluster size as well as to discrete bath degrees of
 freedom.

\subsection{Generalized Luttinger-Ward functional}

Similar to the construction for the nonlocal density-density interaction in Sec. II,
 we first generalize the Luttinger-Ward functional for the Hubbard model with correlated hopping in
 this subsection. For convenience, we define the projected creation
and annihilation operator as
\begin{eqnarray}
    d_{im\sigma}^{\dagger} &=& c_{i \sigma}^{\dagger} P_{i \sigma}^{m}  , \nonumber \\
    d_{im\sigma} &=& c_{i \sigma} P_{i \sigma}^{m}   .     \label{79}
\end{eqnarray}
The statistical action with a source field $\widetilde{G}_{0 ij}^{mn}$ is introduced as

\begin{eqnarray}
    && S\left[ c^{*}, c, \widetilde{G}_{0} \right]    \nonumber \\
    =&& \int_{0}^{\beta} d \tau \int_{0}^{\beta} d \tau^{\prime} \sum_{ij \sigma} \sum_{mn}   \nonumber \\
     & & d_{im \sigma}^{*} \left( \tau \right)
                         \left[ - \left( \widetilde{G}_{0}^{-1} \right)_{ij \sigma}^{mn} \left( \tau - \tau^{\prime }\right)\right]
                           d_{jn \sigma} \left( \tau^{\prime} \right)   \nonumber \\
    & & + \int_{0}^{\beta} d \tau H_{loc} \left( \tau \right)   .    \label{80}
\end{eqnarray}
Equation (80) becomes the corresponding action of the original system when
$\widetilde{G}_{0}= G_{0} $, and
\begin{eqnarray}
    &&   \left( G_{0}^{-1} \right) _{ij \sigma}^{mn} \left( \tau - \tau^{\prime} \right)   \nonumber \\
    &=& \left[ \left( \frac{\partial}{\partial \tau} - \mu\right) \delta_{i j} -t_{i j}^{mn} \right]
                        \delta \left(\tau - \tau^{\prime} \right)    .   \label{81}
\end{eqnarray}
Here, $\widetilde{G}_{0}^{-1}$ should be regarded as the inverse of $\widetilde{G}_{0}$
in the entire space expanded by time, space, spin, and the $m=1,2$ coordinates.
The partition function and the grand potential are
\begin{equation}
    \widetilde{\Xi} \left[ \widetilde{G}_{0} \right] = \int \prod_{i \sigma}
     {\mathcal D} c^{*}_{i \sigma} {\mathcal D} c_{i \sigma}
    e^{-\widetilde{S} [c^{*}, c,\widetilde{G}_{0} ]}      ,           \label{82}
\end{equation}
\begin{equation}
\widetilde{ \Omega} \left[ \widetilde{G}_{0} \right] =
- \frac{1}{\beta} \ln  \widetilde{\Xi}      \left[\widetilde{G}_{0}\right]          .          \label{83}
\end{equation}
The Green's function $\widetilde{G}$ is now defined in a similar way as in Eq. (13),
but based on the operators $d_{im}$ and $d_{im}^{\dagger}$,

\begin{eqnarray}
     & & \widetilde{G}_{ij \alpha}^{mn} \left(\tau - \tau^{\prime} \right)                                   \nonumber \\
     &=&  - \langle T_{\tau} d_{i m\alpha} \left(\tau \right)
              d_{j n \alpha}^{\dagger} \left(\tau^{\prime} \right)\rangle_{\widetilde{S}}   \nonumber \\
     &=& \frac{-1}{\widetilde{\Xi} \left[ \widetilde{G}_{0}\right]}
	      \int \prod_{i \sigma}  {\mathcal D} c^{*}_{i \sigma} {\mathcal D} c_{i \sigma}
                d_{im \alpha} \left(\tau \right) d_{jn \alpha}^{*} \left( \tau^{\prime} \right)
	      e^{-\widetilde{S} [c^{*}, c,\widetilde{ G}_{0} ]}	     .  \nonumber \\
     &&          \label{84}
\end{eqnarray}
From the above definitions, it is easy to establish that
\begin{equation}
        \widetilde{G}_{ij \alpha}^{mn} \left(\tau - \tau^{\prime} \right) = -\beta
	         \frac{\delta \widetilde{\Omega} \left[ \widetilde{G}_{0} \right]}
		 {\delta \left[ \widetilde{G}_{0}^{-1} \right]_{j i \alpha}^{nm} \left(\tau^{\prime} - \tau\right)  }
		      , \label{85}
\end{equation}
which will be abbreviated as
   \begin{equation}
        \widetilde{G} \left[ \widetilde{G}_{0}\right] = -\beta
	         \frac{\delta\widetilde{\Omega} \left[ \widetilde{G}_{0} \right]}
		 {\delta \widetilde{G}_{0}^{-1} }        . \label{86}
\end{equation}
The $\widetilde{G}$ defined above is not the usual electron Green's function.
If we denote the one-particle Green's function as
\begin{equation}
\widetilde{L}_{ij \alpha}\left( \tau - \tau^{\prime}\right)=
- \langle T_{\tau} c_{i \alpha} \left(\tau \right)
              c_{j \alpha}^{\dagger} \left(\tau^{\prime} \right)\rangle_{\widetilde{S}}   , \label{87}
\end{equation}
we have the following relation between $\widetilde{L}$ and $\widetilde{G}$:
\begin{equation}
  \widetilde{L}_{ij \alpha} \left(\tau - \tau^{\prime}\right)
= \sum_{mn} \widetilde{G}_{ij \alpha}^{mn} \left(\tau - \tau^{\prime}\right)  .  \label{88}
\end{equation}
It is noted that similar with the definition of $\{ G, \Pi\}$
in Sec.II, $\widetilde{G}_{0}$ is not the value of
 $\widetilde{G}$ at $H_{loc}=0$. This is because in the action Eq. (80), $\widetilde{G}_{0}$
 is coupled with the occupation-projected operators. Even if $H_{loc}=0$, the action Eq. (80)
  is still a complicated many-particle problem.

For the next step, a Legendre transformation is applied to
$\widetilde{\Omega}[ \widetilde{G}_{0}]$ in a similar way as in the previous section.
 We introduce an intermediate functional $\widetilde{F}$,
\begin{eqnarray}
         \widetilde{F} \left[\widetilde{G}\right] &=& \widetilde{\Omega}
	 \left[\widetilde{G}_{0} \right]  	        - {\text Tr} \left[ \frac{\delta \widetilde{\Omega} \left[ \widetilde{G}_{0}\right]}
			{\delta \widetilde{G}_{0}^{-1}}  \widetilde{G}_{0}^{-1} \right]             \nonumber \\
	       &=& 	\widetilde{\Omega} \left[\widetilde{G}_{0}\right]+ \frac{1}{\beta}
	                   {\text Tr} \left(\widetilde{G} \widetilde{G}_{0}^{-1} \right)  	            ,                \label{89}
\end{eqnarray}
and it follows that
\begin{equation}
         \frac{\delta\widetilde{F} \left[\widetilde{G}\right]}{\delta \widetilde{G}} =
	             \frac{1}{\beta} \widetilde{G}_{0}^{-1}\left[\widetilde{G} \right]      .     \label{90}
\end{equation}
Here, the trace should include the degree of freedom of $m=1,2$.
The generalized self-energy $\widetilde{\Sigma}$ and Luttinger-Ward functional
$\widetilde{\Phi}[\widetilde{G} ]$ are defined in a similar fashion,
\begin{equation}
\widetilde{\Sigma}=\widetilde{G}_{0}^{-1}-\widetilde{G}^{-1}  , \label{91}
\end{equation}
\begin{equation}
       \widetilde{\Phi} \left[\widetilde{G} \right]= \beta \widetilde{F} \left[\widetilde{G}\right]
                 - {\text Tr} \ln \widetilde{G}     .                  \label{92}
\end{equation}
Note that the parameter $\alpha$ introduced in the previous section is now selected as $\alpha=-1$,
to be consistent with the pure fermion field components of $d_{im\sigma}$.
We get a formula formally same as the one of the original Luttinger-Ward functional,
\begin{equation}
        \frac{\delta\widetilde{\Phi} \left[\widetilde{G}\right] }{ \delta \widetilde{G}}
	   = \widetilde{\Sigma}               ,     \label{93}
\end{equation}
and
\begin{eqnarray}
       \beta \widetilde{\Omega}\left[\widetilde{G}\right] &=&  \widetilde{\Phi} \left[\widetilde{G} \right]
	+ {\text Tr} \ln \widetilde{G} - {\text Tr} \left( \widetilde{G} \widetilde{G}_{0}^{-1} \right)     .      \label{94}
 \end{eqnarray}
The difference here is that the $\widetilde{G}$, $\widetilde{G}_{0}$ and $\Sigma$ are matrices
in the expanded space including index $m=1,2$.
The Luttinger-Ward expression of the functional of $\widetilde{\Omega}$ is obtained
immediately by simply replacing the $\widetilde{G}_{0}$ with its physical value $G_{0}$, and we get
\begin{eqnarray}
       \beta \widetilde{\Omega}_{LW}\left[\widetilde{G} \right] &=&  \widetilde{\Phi} \left[\widetilde{G}\right]
	+ {\text Tr} \ln \widetilde{G} - {\text Tr} \left( \widetilde{G} G_{0}^{-1} \right)           ,      \label{95}
 \end{eqnarray}
with the following stationary properties
\begin{eqnarray}
  && \frac{\delta \widetilde{\Omega}_{LW} \left[\widetilde{G}\right]}
  {\delta \widetilde{G}} |_{\widetilde{G}=G}=0  , \nonumber \\
&&\widetilde{\Omega}_{LW} \left[\widetilde{G}=G \right]= \Omega .   \label{96}
\end{eqnarray}

\subsection{EVCA for correlated hopping }
In this subsection, we continue to construct the EVCA equation for the correlated hopping system.
Doing Legendre transformation for $\widetilde{\Phi} [ \widetilde{G} ]$, we get
\begin{eqnarray}
         \widetilde{A} \left[\widetilde{\Sigma} \right] &=& \widetilde{\Phi}
	 \left[\widetilde{G} \right]
	  - {\text Tr} \left[ \frac{\delta \widetilde{\Phi} \left[ \widetilde{G} \right]}
		{\delta \widetilde{G}}  \widetilde{G} \right]             \nonumber \\
	     &=& 	\widetilde{\Phi} \left[\widetilde{G} \right]
        - {\text Tr} \left( \widetilde{\Sigma} \widetilde{G} \right)        ,           \label{97}
\end{eqnarray}
and
\begin{equation}
         \frac{\delta \widetilde{A} \left[\widetilde{\Sigma} \right]}{\delta \widetilde{\Sigma}} =
	            -\widetilde{G}\left[\widetilde{\Sigma}  \right]   .     \label{98}
\end{equation}
The grand potential $\widetilde{\Omega}$ as a functional of the generalized self-energy
$\widetilde{\Sigma}$ reads
\begin{equation}
    \beta \widetilde{\Omega}\left[\widetilde{\Sigma} \right]
    = \widetilde{A} \left[\widetilde{\Sigma} \right]
            -{\text Tr} \ln \left( \widetilde{G}_{0}^{-1} -\widetilde{\Sigma}\right)    .              \label{99}
\end{equation}
The grand potential functional having the desired stationary properties is defined as
\begin{equation}
    \beta \widetilde{\Omega}_{EVCA}\left[\widetilde{\Sigma} \right]
    = \widetilde{A} \left[\widetilde{\Sigma} \right]
            -{\text Tr} \ln \left( G_{0}^{-1} -\widetilde{\Sigma}\right)   .              \label{100}
\end{equation}
It is obtained by replacing $\widetilde{G}_{0}^{-1}$ in Eq. (99) with its physical value
$G_{0}^{-1}$ in Eq. (81). Using Eq. (99) to eliminate $\widetilde{A}$ in Eq. (100),
we get
\begin{eqnarray}
&&     \widetilde{\Omega}_{EVCA}\left[\widetilde{\Sigma} \right]  \nonumber \\
  &=& \widetilde{\Omega}\left[\widetilde{\Sigma} \right]
    +\frac{1}{\beta} {\text Tr} \ln \left( \widetilde{G}_{0}^{-1} - \widetilde{\Sigma}\right)
      -\frac{1}{\beta} {\text Tr} \ln \left( G_{0}^{-1} -\widetilde{\Sigma}\right)   .      \nonumber\\
   &&           \label{101}
\end{eqnarray}

Now we introduce the reference system. As usual, the reference system is defined
on a clusterized lattice with the same $H_{loc}$ part as the original Hamiltonian (78).
The action reads

\begin{eqnarray}
    && S_{ref} \left[ c^{*}, c, \widetilde{G}_{0c} \right]    \nonumber \\
    =&& \int_{0}^{\beta} d \tau \int_{0}^{\beta} d \tau^{\prime} \sum_{ij \sigma} \sum_{mn}   \nonumber \\
     & & d_{im \sigma}^{*} \left( \tau \right)
                         \left[ - \left( \widetilde{G}_{0c}^{-1} \right)_{ij \sigma}^{mn} \left( \tau - \tau^{\prime }\right)\right]
                           d_{jn \sigma} \left( \tau^{\prime} \right)   \nonumber \\
    & & + \int_{0}^{\beta} d \tau \, H_{loc} \left( \tau \right)     ,    \label{102}
\end{eqnarray}
where the Weiss field $\widetilde{G}_{0c}^{-1}$ is block diagonal in the spatial coordinate.
The generalized self-energy functional $\widetilde{\Omega}_{EVCA}$ for the reference system
can be obtained along the same line, except that now the generalized self-energy
$\widetilde{\Sigma}$ should be replaced with $\widetilde{\Sigma}_{c}$. By varying
the parameters of the reference system (102), $\widetilde{\Sigma}_{c}$ can be varied through out
its space. Here we have made use of the fact that the same functional form $\widetilde{\Omega}_{EVCA}
[ \widetilde{\Sigma} ]$ applies both for the original system and for the reference system.
We then obtain
\begin{eqnarray}
   \widetilde{\Omega}_{EVCA}\left[\widetilde{\Sigma}_{c} \right]
  &=& \widetilde{\Omega}\left[\widetilde{\Sigma}_{c} \right]
    +\frac{1}{\beta} {\text Tr} \ln \left( \widetilde{G}_{0c}^{-1} - \widetilde{\Sigma}_{c}\right)   \nonumber \\
    &&  -\frac{1}{\beta} {\text Tr} \ln \left( G_{0}^{-1} -\widetilde{\Sigma}_{c}\right)  .              \label{103}
\end{eqnarray}
Using the same principle of constructing EVCA as before,
We establish the EVCA equation for a continuous reference system,
\begin{equation}
  \frac{\delta \widetilde{\Omega}_{EVCA} \left[\widetilde{\Sigma}_{c}\right]}
  {\delta \widetilde{\Sigma}_{c}}  |_{\widetilde{\Sigma}_{c}=\Sigma_{app} }   =0
        .      \label{104}
\end{equation}
For a discrete reference system where the Weiss field $\widetilde{G}_{0c}$
is parametrized by finite number of parameters $\widetilde{t}$, the EVCA equation
reads

\begin{equation}
   \frac{\delta \widetilde{\Omega}_{EVCA}  \left[ \widetilde{\Sigma}_{c}\left( \widetilde{t} \right) \right]}
  {\delta \widetilde{t} }  |_{\widetilde{t}=t_{app}}  =0             ,      \label{105}
\end{equation}
and $\Sigma_{app}=\widetilde{\Sigma}_{c}\left( \widetilde{t} = t_{app} \right)$.
Introducing Eq. (103) into Eq.(104) and (105), we get the continuous version of EVCA,
\begin{eqnarray}
  &&   \widetilde{G}_{c} = \left|  \left( G_{0}^{-1} - \widetilde{\Sigma}_{c}  \right) \right|  \nonumber \\
  &&  \widetilde{\Sigma}_{c}=  \widetilde{G}_{0c}^{-1} - \widetilde{G}_{c}^{-1}
               .       \label{106}
\end{eqnarray}
Here the cluster index $I$ is dropped.
This set of equations has the same form as the VCA equation for the pure Hubbard model,
but with different definitions of $\widetilde{G}_{c}$. Here $\widetilde{G}_{c}$ is the two-particle
Green's function instead of the usual one-particle Green's function.
In the case of one-impurity reference system, Eq. (106) reduces to the
equation of DMFTCH proposed by  Shvaika, \cite{Shvaika1} which is proved to be exact
in infinite spatial dimension. \cite{Schiller2} In this sense, Eq. (106) is a generalization
of DMFTCH to cluster reference system.
For the reference system with discrete degrees of freedom, Eq. (105) produces for each cluster
index $I$,
\begin{equation}
    {\text Tr} \left[ \widetilde{G}_{c} - \left|  \left( G_{0}^{-1} - \widetilde{\Sigma}_{c}  \right)^{-1} \right| \right]
            \frac{\delta \widetilde{\Sigma}_{c}\left( \widetilde{t} \right) }{\delta \widetilde{t}}
	     =0        .      \label{107}
\end{equation}
This discrete version of EVCA equation allows for more flexible and efficient
 selection of the reference system. It is actually a more general form, since in the continuous
 limit, the continuous EVCA equation Eq. (106) is recovered.

\section{Translation-Invariant VCA and EVCA }

In this section, we put forward the VCA and EVCA equations
that are based on a cluster reference system with periodic boundary conditions.
When using a reference system with continuous degrees of freedom, the VCA
will recover the DCA of Jarrell {\it et al.}, and the EVCA will become
an extension of the DCA including nonlocal interactions.

The DCA and CDMFT are
developed independently to extend the original single-impurity DMFT formalism
to finite clusters. The CDMFT is designed on clusters
 with open boundary conditions, while the DCA adopts periodic boundary conditions
 for the cluster. Both formalisms recover the DMFT in the one-impurity
case, and become exact in the limit of large cluster size. However, with increasing cluster size,
the two methods have apparently different convergence properties. Recently, detailed comparisons
between the performance of DCA and CDMFT have been made.\cite{Biroli1, Aryanpour1} It was pointed out
that the real-space formula of DCA can be obtained by replacing the bare electron dispersion $\epsilon({\bf k})$
in the CDMFT equations with a different one $\epsilon^{cyc}({\bf K}, {\bf K}_c)$. This means
that the formula of CDMFT can be changed into that of DCA by simply replacing
the bare $G_{0 CDMFT}^{-1}$ in that theory with the corresponding $G_{0 DCA} ^{-1}$,
 and vice versa.

 The VCA (Ref. 24) and our construction of EVCA in
 Secs. II and III  all make use of clusters with open boundary conditions. They can recover
 the CDMFT and EDMFT (or DMFTCH), respectively, in the case of continuous bath degrees of freedom.
 However, since these theories give out self-energies which
 are not translation invariant, some kind of estimator should be used to calculate the
 lattice self-energies from the cluster ones. Here, we construct a different version of VCA
 such that it has a momentum-space representation similar to DCA. In the continuous limit,
 it recovers the formalism of DCA. Similarly, the EVCA formalism developed in previous
 sections can also be adapted to using periodic boundary conditions for the
 clusters of the reference system. In the continuous limit, it leads to an extension of DCA formula
 to treat nonlocal interactions (or correlated hopping).

Our discussion in the following is for a $D$-dimensional hypercubic lattice. Other
 lattice structure can be discussed similarly. The number of sites and site interval
 for the $i$th dimension are denoted as $N_{i}$
and $a_{i}$, respectively. We divide the whole lattice into geometrically identical
 clusters of size $\left( L_{1}a_{1}, L_{2}a_{2}, ..., L_{D}a_{D} \right)$. The number
 of total clusters is $N/L=\prod_{i=1}^{D}{N_{i}/L{i}}$. The site vector of a cluster is denoted
 as ${\bf  l} = \left( l_1 a_1, l_2 a_2, ..., l_D a_D \right)$, with $l_i = 0,...,L_i-1$.
We define the super-lattice site vector as ${\bf r}=\left( r_1 L_1 a_1, r_2 L_2 a_2, ..., r_D L_D a_D \right)$,
with $r_i = 0,...,N_i /L_i -1 $.
  Any lattice vector
 ${\bf  n}= \left( n_1 a_1, n_2 a_2, ..., n_D a_D \right)$ with $n_i= 0, 1, ..., N_i-1$ is uniquely
 expressed as the summation of a super lattice site vector and a cluster site vector,
 ${\bf n} ={\bf l} + {\bf r}$. In the first Brillouin zone of the three lattices, the original lattice,
 the cluster lattice, and the super lattice, their corresponding inverse vectors are denoted as
 ${\bf k }$, ${\bf K}_{c}$, and ${\bf K}$, respectively. We have

 \begin{eqnarray}
   && {\bf k}=\left( \frac{2 \pi m_1}{ N_1 a_1},  ..., \frac{2 \pi m_D}{ N_D a_D} \right) \,\,\,\,\,\,
                            \left( m_i = 0, 1, ..., N_i -1 \right),  \nonumber \\
  &&  {\bf K}_{c}=\left( \frac{2 \pi p_1}{ L_1 a_1},  ..., \frac{2 \pi p_D}{ L_D a_D} \right)  \, \,\,\, \,\,
                           \left( p_i = 0, 1, ..., L_i -1   \right),  \nonumber  \\
 &&  {\bf K}=\left( \frac{2 \pi q_1}{ N_1 a_1},  ..., \frac{2 \pi q_D}{ N_D a_D} \right)    \, \,\, \, \,\,
                           \left( q_i = 0, 1, ..., N_i/L_i -1  \right)  .    \nonumber \\
  &&      \label{108}
\end{eqnarray}
Similarly, any lattice momentum ${\bf k}$ is uniquely expressed by ${\bf k}={\bf K}_{c}+ {\bf K}$.
The orthogonal and complete relations read
\begin{eqnarray}
  &&\frac{1}{N} \sum_{{\bf k}} e^{i {\bf k} \cdot \left( {\bf n}_1 - {\bf n}_2 \right)} = \delta_{ {\bf n}_1, {\bf n}_2}  ,  \nonumber \\
  && \frac{1}{L} \sum_{{\bf K}_c} e^{i {\bf K}_c \cdot \left( {\bf l}_1 - {\bf l}_2 \right) } = \delta_{{\bf l}_1, {\bf l}_2}  ,  \nonumber \\
  && \frac{L}{N} \sum_{\bf K} e^{i {\bf K} \cdot \left( {\bf r}_1 - {\bf r}_2 \right) } = \delta_{ {\bf r}_1, {\bf r}_2}  ,  \label{109}
\end{eqnarray}
and
\begin{eqnarray}
  &&\frac{1}{N} \sum_{\bf n} e^{i {\bf n} \cdot \left( {\bf k}_1 - {\bf k}_2 \right) } = \delta_{ {\bf k}_1, {\bf k}_2}  ,  \nonumber \\
  && \frac{1}{L} \sum_{\bf l} e^{i {\bf l}\cdot  \left( {\bf K}_{c1} - {\bf K}_{c2} \right) }
                          = \delta_{{\bf K}_{c1}, {\bf K}_{c2}}  ,  \nonumber \\
  && \frac{L}{N} \sum_{\bf r} e^{i {\bf r} \cdot \left( {\bf K}_1 - {\bf K}_2 \right) } = \delta_{ {\bf K}_1, {\bf K}_2}  .  \label{110}
\end{eqnarray}
Here, we consider translation-invariant $t_{{\bf n}_1 {\bf n}_2 }$
and assume that it has a periodic boundary condition on the $D$-dimensional lattice. Also, only
translation-invariant solutions of $\widetilde{G}_{c}$ and $\widetilde{\Sigma}_{c}$ are considered.
For lattice symmetry-broken solutions, one should take care of the sublattice indices.
 The VCA equation is formally the same as the EVCA equation Eq. (106) for the correlated
hopping,
\begin{eqnarray}
  &&   \widetilde{G}_{c} = \left|  \left( G_{0}^{-1} - \widetilde{\Sigma}_{c}  \right)^{-1} \right| ,  \nonumber \\
  &&  \widetilde{\Sigma}_{c}=  \widetilde{G}_{0c}^{-1} - \widetilde{G}_{c}^{-1}
              .       \label{111}
\end{eqnarray}
Here the Green's function is the usual one-particle Green's function, and $G_{0}^{-1}$ in the frequency axis is
\begin{equation}
 G_{0 {\bf n}_1 {\bf n}_2 }^{-1} \left( i \omega_n \right) = \left( i \omega_n + \mu \right) \delta_{{\bf n}_1 {\bf n}_2}
 +t_{{\bf n}_1 {\bf n}_2 }      .          \label{112}
\end{equation}
In the spatial coordinate, $\widetilde{G}_{c}$ and $\widetilde{\Sigma}_{c}$ are block diagonal
and $G_{0}^{-1}$ is not since the hopping matrix of the original Hamiltonian $t_{{\bf n}_1 {\bf n}_2 }$
is not limited within the cluster.
We write $[G_{0}^{-1} ]_{{\bf n}_1 {\bf n}_2}$ as $[G_{0}^{-1} ]_{{\bf l}_1 {\bf l}_2}^{{\bf r}_1 {\bf r}_2}$,
where the upper index denotes the superlattice site vector,
and the lower one denotes the cluster site vector. The Fourier transformation to the super lattice index
gives
\begin{eqnarray}
 [G_{0}^{-1}]_{{\bf l}_1 {\bf l}_2} \left( {\bf K}\right) &=& \sum_{{\bf r}_{1}-{\bf r}_{2} }
  [G_{0}^{-1} ]_{{\bf l}_1 {\bf l}_2}^{{\bf r}_1 {\bf r}_2} e^{-i {\bf K} \cdot \left( {\bf r}_{1}-{\bf r}_{2}\right)}   \nonumber \\
    &=&\left( i \omega_n + \mu \right) \delta_{{\bf l}_1 {\bf l}_2}
    - \hat{\epsilon}_{{\bf l}_1 {\bf l}_2} \left( {\bf K}\right)   . \label{113}
\end{eqnarray}
The matrix $\hat{ \epsilon } \left( {\bf K}\right)$ is defined as
\begin{equation}
\hat{\epsilon}_{{\bf l}_1 {\bf l}_2} \left( {\bf K}\right) = \frac{1}{L} \sum_{{\bf K}_{c}}
        \epsilon \left( {\bf K}+ {\bf K}_{c} \right) e^{ i \left( {\bf K}+ {\bf K}_{c} \right) \cdot \left( {\bf l}_1- {\bf l}_2 \right) }
	 ,   \label{114}
\end{equation}
where $\epsilon \left( {\bf k}= {\bf K}+ {\bf K}_{c} \right)= \sum_{{\bf n}_1 - {\bf n}_2 }
  - t_{{\bf n}_1 {\bf n}_2 } e^{-i {\bf k} \left( {\bf n}_1 - {\bf n}_2 \right) }$ is the dispersion of free electrons.

  The same Fourier transformation on $\widetilde{G}_{c}$ and  $\widetilde{\Sigma}_{c}$ give out
\begin{eqnarray}
 && \widetilde{G}_{c {\bf l}_1 {\bf l}_2} \left( {\bf K}\right) = \sum_{{\bf r}_{1}-{\bf r}_{2} }
   \widetilde{G}_{c {\bf l}_1 {\bf l}_2}^{{\bf r}_1 {\bf r}_2} e^{-i {\bf K} \cdot \left( {\bf r}_{1}-{\bf r}_{2}\right)}
    = \widetilde{G}_{{c {\bf l}_1 {\bf l}_2}}^{I}     ,                \nonumber \\
&& \widetilde{\Sigma}_{c {\bf l}_1 {\bf l}_2} \left( {\bf K}\right) = \sum_{{\bf r}_{1}-{\bf r}_{2} }
   \widetilde{\Sigma}_{c {\bf l}_1 {\bf l}_2}^{{\bf r}_1 {\bf r}_2} e^{-i {\bf K} \cdot \left( {\bf r}_{1}-{\bf r}_{2}\right)}
    = \widetilde{\Sigma}_{ {c {\bf l}_1 {\bf l}_2}}^{I}            .   \nonumber \\
  &&     \label{115}
\end{eqnarray}
Here $\widetilde{G}_{c}^{I}$ and $\widetilde{\Sigma}_{c}^{I}$ are matrices defined on one of the clusters, which
is actually independent of the cluster index $I$.
We have
\begin{eqnarray}
   && \left | \left( G_{0}^{-1} - \widetilde{\Sigma}_{c} \right)^{-1} \right| ^{ I }_{{\bf l}_1 {\bf l}_2}  \nonumber \\
  &=&  \frac{L}{N} \sum_{{\bf K}} \left[  G_{0}^{-1} \left({\bf K}\right) -\widetilde{\Sigma}_{c}
               \left( {\bf K} \right) \right]^{-1}_{{\bf l}_1 {\bf l}_2}          \nonumber \\
  &=&\frac{L}{N} \sum_{{\bf K}} \left[  i \omega_{n} + \mu - \hat{\epsilon} ({\bf K}) - \widetilde{\Sigma}_{c}^{I}
                \right]^{-1}_{{\bf l}_1 {\bf l}_2}          . \label{116}
\end{eqnarray}
Finally the VCA equation Eq. (111) is simplified to

\begin{eqnarray}
  && \widetilde{G}_{c}^{I} \left( i \omega_n \right) =
  \frac{L}{N} \sum_{{\bf K}} \left[  i \omega_{n} + \mu - \hat{\epsilon} \left({\bf K} \right) - \widetilde{\Sigma}_{c}^{I}
               \left( i \omega_n \right) \right] ^{-1}     ,   \nonumber \\
  && \widetilde{\Sigma}_{c}^{I}\left( i \omega_n \right) =   \widetilde{G}_{0c}^{I \, -1}
  \left( i \omega_n \right)- \widetilde{G}_{c}^{I \, -1} \left( i \omega_n \right)
		   .  \label{117}
\end{eqnarray}
Note that ${\bf K}$ summation is only carried out in the reduced Brillouin zone of the super lattice.
Equation (117) is exactly the CDMFT formula presented in Ref. 13.
The real-space formulation of DCA in Ref. 13 shows that when $\hat{\epsilon} \left({\bf K} \right)$
in Eq. (117) is replaced by
a  $\hat{\epsilon}^{cyc}\left({\bf K} \right) $, Eq. (117) becomes the DCA equation.
$\hat{\epsilon}^{cyc}\left({\bf K} \right) $ is defined as
\begin{equation}
 \hat{\epsilon}^{cyc}_{{\bf l}_1 {\bf l}_2}\left({\bf K} \right)=\frac{1}{L} \sum_{{\bf K}_{c}}
        \epsilon \left( {\bf K}+ {\bf K}_{c} \right) e^{ i  {\bf K}_{c} \cdot \left( {\bf l}_1- {\bf l}_2 \right) }
	 .   \label{118}
\end{equation}
The periodic boundary condition of $\hat{\epsilon}^{cyc}_{{\bf l}_1 {\bf l}_2}$ makes it possible to
do further Fourier transformation on Eq. (117) over the cluster site index, and we obtain the
original DCA equation established on the cluster ${\bf K}_c$ points:
\begin{eqnarray}
  && \widetilde{G}_{c} \left( {\bf K}_c \right) =
  \frac{L}{N} \sum_{{\bf K}} \left[  i \omega_{n} + \mu - \epsilon \left({\bf K}+{\bf K}_c \right) - \widetilde{\Sigma}_{c}
               \left( {\bf K}_{c}\right) \right] ^{-1}    ,    \nonumber \\
  && \widetilde{\Sigma}_{c}\left( {\bf K}_{c} \right) =   \widetilde{G}_{0c}\left( {\bf K}_{c} \right)^{-1}
  - \widetilde{G}_{c}\left( {\bf K}_{c} \right) ^{-1} 		   .  \label{119}
\end{eqnarray}
DCA can produce a translation-invariant cluster self-energy, which can be directly identified as
the lattice self-energy, in contrast to CDMFT.

We are interested to see what $\hat{\epsilon}^{cyc}$ looks like in real space. It is found that
the corresponding real-space hopping matrix element of $\hat{\epsilon}^{cyc}$ is
\begin{equation}
  [ t^{cyc} ] ^{{\bf r}_1 {\bf r}_2}_{{\bf l}_1 {\bf l}_2}=- \frac{1}{N} \sum_{{\bf K}, {\bf K}_c}
   \epsilon \left({\bf K}+{\bf K}_c \right)e^{i {\bf K} \cdot \left( {\bf r}_1- {\bf r}_2 \right)+
    i {\bf K}_{c} \left( {\bf l}_1- {\bf l}_2 \right) }   ,    \label{120}
\end{equation}
The full Fourier transformation of  $[t^{cyc}]^{{\bf r}_1 {\bf r}_2}_{{\bf l}_1 {\bf l}_2}$ gives an electron
dispersion
\begin{equation}
  \epsilon^{cyc} \left( {\bf K}, {\bf K}_{c} \right)= \frac{1}{L} \sum_{ {\bf K}_c^{\prime}} \sum_{{\bf l}_1- {\bf l}_2}
   \epsilon \left({\bf K}+{\bf K}_c^{\prime} \right)e^{i \left( {\bf K}_{c}^{\prime} - {\bf K}_{c} -{\bf K}\right) \cdot
        \left( {\bf l}_1- {\bf l}_2 \right) }   .   \label{121}
\end{equation}
The real-space formula of DCA can be expressed in our notation as
\begin{eqnarray}
  &&   \widetilde{G}_{c} = \left|  \left( G_{0 DCA}^{-1} - \widetilde{\Sigma}_{c}  \right)^{-1} \right|  , \nonumber \\
  &&  \widetilde{\Sigma}_{c}=  \widetilde{G}_{0c}^{-1} - \widetilde{G}_{c}^{-1}
              ,       \label{122}
\end{eqnarray}
where
\begin{equation}
 [G_{0 DCA} ^{-1}]_{ {\bf n}_1 {\bf n}_2 } \left( i \omega_n \right) = \left( i \omega_n + \mu \right) \delta_{{\bf n}_1 {\bf n}_2}
 +t^{cyc}_{{\bf n}_1 {\bf n}_2 }      .          \label{123}
\end{equation}
This can be easily verified since the Fourier transformation to Eq. (122) over the superlattice site vector
will lead to the real-space formulation of DCA. Comparison of Eq. (111) and Eq. (122) shows that
DCA and CDMFT can be described by a unified equation with different bare electron dispersion.
One is $\epsilon({\bf k})$ for CDMFT; the other is $\epsilon^{cyc} \left( {\bf K}, {\bf K}_{c} \right)$ for DCA.
The corresponding bare Green's function entering the theory is $G_{0}$ and $G_{0 DCA}$, respectively.
Because $\epsilon \left( {\bf k}={\bf K}_{c} \right) = \epsilon^{cyc} \left( {\bf K}, {\bf K}_{c} \right)$,
when cluster size $L$ increases, the number of $K_c$ points increases.
$\epsilon^{cyc} \left( {\bf K}, {\bf K}_{c} \right)$ will tend to $\epsilon \left( {\bf k} \right)$
uniformly in the ${\bf k}$ space. This shows the equivalence of CDMFT and DCA in the large-cluster limit.

The merit of the real-space formula of DCA is that it not only has a translation invariant form as in the
${\bf k}$-space formula,
 but also allows for the study of the phases with reduced translation invariance or system without translation invariance,
  such as an antiferromagnetic phase or disordered system.
 With the above discussion, it is straightforward to put down the VCA equations for a general reference system while still
 keeping a translation invariant form.
 This is done simply by using $G_{0 DCA} ^{-1}$ in the original VCA equations. We get
 \begin{equation}
    {\text Tr} \left[ \widetilde{G}_{c} - \left|  \left( G_{0 DCA}^{-1} - \widetilde{\Sigma}_{c}  \right)^{-1} \right| \right]
            \frac{\delta \widetilde{\Sigma}_{c}\left( \widetilde{t} \right) }{\delta \widetilde{t}}
	     =0       .      \label{124}
\end{equation}
This is the DCA equation adapted for discrete reference system. In the continuous bath limit, it recovers the original
DCA equation.\cite{Hettler1} DCA has already been
implemented with various cluster solvers.\cite{Maier1, Aryanpour1, Maier2} With Eq. (124), it is now
possible to study the discrete version of DCA on reference systems with smaller cluster Hilbert space
using numerical methods such as exact diagonalization. And the flexibility
and efficiency of the VCA is thus combined with the merit of translation-invariant self-energy of DCA.

For the EVCA, a translation invariant form can also be obtained along the same line. For the case of the
density-density interaction, we use the modified hopping matrix $t^{cyc}_{ {\bf n}_1 {\bf n}_2}$ Eq. (120),
and define the interacting parameters $V^{cyc}$ as
\begin{equation}
  [ V^{cyc} ] ^{{\bf r}_1 {\bf r}_2}_{{\bf l}_1 {\bf l}_2}=- \frac{1}{N} \sum_{{\bf K}, {\bf K}_c}
   V \left({\bf K}+{\bf K}_c \right)e^{i {\bf K} \cdot \left( {\bf r}_1- {\bf r}_2 \right)+
    i {\bf K}_{c} \left( {\bf l}_1- {\bf l}_2 \right) }    .    \label{125}
\end{equation}
Replacing $G_{0}^{-1}$ and $\Pi_{0}^{-1}$ in Eqs. (59) and (60) with the corresponding DCA version
Eq. (123) and
\begin{equation}
 [\Pi_{0 DCA} ^{-1}]_{ {\bf n}_1 {\bf n}_2 }  = - V^{cyc}_{{\bf n}_1 {\bf n}_2 }  ,          \label{126}
\end{equation}
we conveniently get the general real space formulas for EVCA with translation invariance,
\begin{eqnarray}
  &&   {\text Tr} \left[ \widetilde{G}_{c} - \left|  \left( G_{0 DCA}^{-1} - \widetilde{\Sigma}_{c}  \right)^{-1} \right| \right]
            \frac{\delta \widetilde{\Sigma}_{c}\left( \widetilde{t}, \widetilde{V} \right) }{\delta \widetilde{t}}  \nonumber \\
  &&  +{\text Tr} \left[ \widetilde{\Pi}_{c} - \left|  \left( \Pi_{0 DCA}^{-1} - \widetilde{\Gamma}_{c}  \right)^{-1} \right| \right]
             \frac{\delta \widetilde{\Gamma}_{c}\left( \widetilde{t}, \widetilde{V} \right) }{\delta \widetilde{t}}
	     =0        ,   \nonumber \\
	&&        \label{127}
\end{eqnarray}
\begin{eqnarray}
 &&   {\text Tr} \left[ \widetilde{G}_{c} - \left|  \left( G_{0 DCA}^{-1} - \widetilde{\Sigma}_{c}  \right)^{-1} \right| \right]
            \frac{\delta \widetilde{\Sigma}_{c}\left( \widetilde{t}, \widetilde{V} \right) }{\delta \widetilde{V}}  \nonumber \\
  &&  +{\text Tr} \left[ \widetilde{\Pi}_{c} - \left|  \left( \Pi_{0 DCA}^{-1} - \widetilde{\Gamma}_{c}  \right)^{-1} \right| \right]
             \frac{\delta \widetilde{\Gamma}_{c}\left( \widetilde{t}, \widetilde{V} \right) }{\delta \widetilde{V}}
	     =0        .      \nonumber \\
  &&	          \label{128}
\end{eqnarray}
Another version of EVCA decouples the fermion and boson contributions to the grand potential.
Its corresponding translation invariant form reads
\begin{eqnarray}
  &&   {\text Tr} \left[ \widetilde{G}_{c} - \left|  \left( G_{0 DCA}^{-1} - \widetilde{\Sigma}_{c}  \right)^{-1} \right| \right]
            \frac{\delta \widetilde{\Sigma}_{c}\left( \widetilde{t}, \widetilde{V} \right) }{\delta \widetilde{t}} =0  , \nonumber \\
  &&  {\text Tr} \left[ \widetilde{\Pi}_{c} - \left|  \left( \Pi_{0 DCA}^{-1} - \widetilde{\Gamma}_{c}  \right)^{-1} \right| \right]
             \frac{\delta \widetilde{\Gamma}_{c}\left( \widetilde{t}, \widetilde{V} \right) }{\delta \widetilde{V}}
	     =0   . \nonumber \\
  &&        \label{129}
\end{eqnarray}
In the continuous bath limit, the above EVCA theories become the extension of DCA to include nonlocal
interactions, and can be denoted as EDMFT$+$DCA.

Similarly, the EVCA for correlated hopping can also be implemented in a translation invariant way. We simply
replace the bare Green's function in Eq. (107) with the DCA version, and the obtained EVCA equation reads
\begin{equation}
    {\text Tr} \left[ \widetilde{G}_{c} - \left|  \left( G_{0 DCA}^{-1} - \widetilde{\Sigma}_{c}  \right)^{-1} \right| \right]
            \frac{\delta \widetilde{\Sigma}_{c}\left( \widetilde{t} \right) }{\delta \widetilde{t}}
	     =0        ,      \label{130}
\end{equation}
where $G_{0 DCA}^{-1} \left( i \omega_n \right)$ can be described by Eqs. (120) and (123),
but with the $t^{cyc}$ and $\epsilon \left( {\bf k}\right)$ being $2 \times 2$ matrices in the space of single occupation
and nonoccupation.

\section{Discussion}

In this section, several issues about EVCA are discussed. We first discuss the EVCA equations for the ordered
phase. Then, the EVCA is constructed after carrying out a Hubbard-Stratonovich transformation on the original system.
Comparison with previous EVCA equations shows that the Hubbard-Stratonovich transformation plays a nontrivial role here.
Using the same method, the EVCA for the Heisenberg model, a correlated spin system is obtained straightforwardly.
 The problems with EVCA for bosonic systems are discussed. Finally, we point out that our construction applies
 equally well to classical models, such as the Ising model.

\subsection{EVCA for ordered phase}
When the system is in the ordered phase induced by the corresponding nonlocal interactions,
the implementation of EVCA need to be modified according to broken symmetry.
This issue has been discussed thouroughly in the context of EDMFT.\cite{Sun1,Chitra1,Smith1}
 Here, a similar strategy can be taken to handle the ordered phase in EVCA.
EVCA equations developed in Secs. II and III, such as Eqs. (59), (60), and (63), are still valid.
The Hartree contribution should be singled out by using the normal ordered operator in the original Hamiltonian.
 However, care should be taken when we write down the reference Hamiltonian and
evaluate the trace terms in those equations. In the case of a bipartite charge-ordered phase, for example,
the Weiss fields appearing in the action of the reference system should obey the two-sublattice symmetry
 of the ordered phase.
The trace terms in Eqs. (59), (60), and (63) should also be evaluated on the two sublattices.
These will lead to different final expressions for the $\Omega_{EVCA}$ functional and
EVCA equations from the symmetry-unbroken case. Detailed expression will be discussed elsewhere
in combination with model studies.

\subsection{The role of Hubbard-Stratonovich transformation}

Up to now, the EVCA equations discussed in this paper are constructed directly for the original electron
Hamiltonian. One can also first carry out a Hubbard-Stratonovich (HS) transformation on the nonlocal
interaction terms in the Hamiltonian, and construct the EVCA equations for the resulting effective electron-phonon
Hamiltonian. This latter approach is widely used in EDMFT studies since it facilitates the employment of the
Monte Carlo algorithm. It has been claimed that the two approaches are equivalent to each other.\cite{Sun1} Here
taking the extended Hubbard model Eq. (3) as an example, we study the role of HS transformation.
 It is found that in general the two approaches are nonequivalent.
For Hamiltonian (3), the direct construction starts from the action (11),
\begin{eqnarray}
  & & \widetilde{S}_{A} [ c^{*}, c, \widetilde{G}_{0}, \widetilde{\Pi}_{0} ]        \nonumber \\
   &=&  \int_{0}^{\beta} d \tau \int_{0}^{\beta} d \tau^{\prime} \sum_{i j \sigma} c_{i \sigma}^{*} \left(\tau \right)
          [ - \widetilde{G}_{0}^{-1} ]_{i j \sigma} \left( \tau - \tau^{\prime} \right) c_{j \sigma} \left( \tau^{\prime} \right)          \nonumber \\
   &+&   \int_{0}^{\beta} d \tau \int_{0}^{\beta} d \tau^{\prime} \sum_{i j} : n_{i} \left(\tau \right) :
          [ - \widetilde{\Pi}_{0}^{-1} ]_{i j} \left( \tau - \tau^{\prime} \right) : n_{j} \left( \tau^{\prime} \right) :           \nonumber \\
   &+&  \int_{0}^{\beta} H^{\prime}_{loc A} \left(\tau \right) d \tau                      ,          \label{131}
\end{eqnarray}
where
\begin{eqnarray}
H^{\prime}_{loc A}\left( \tau \right) &=&
H_{loc}\left( \tau \right)+ 2\sum_{ij} n_{i}\left( \tau \right) \left[ -\Pi_{0}^{-1}\right]_{ij} \langle n_{j} \rangle \nonumber \\
                &+& \sum_{ij} \Pi_{0 ij}^{-1} \langle n_{i} \rangle \langle n_{j} \rangle    .      \label{132}
\end{eqnarray}
The corresponding EVCA equations obtained for the continuous reference system read [Eq. (55)]
\begin{eqnarray}
  &&   \widetilde{G}_{c} = \left|  \left( G_{0}^{-1} - \widetilde{\Sigma}_{c}  \right)^{-1} \right|  , \nonumber \\
  &&   \widetilde{\Pi}_{c} = - \alpha \left|  \left( \Pi_{0}^{-1} - \widetilde{\Gamma}_{c}  \right)^{-1} \right|  ,  \nonumber \\
  &&  \widetilde{\Sigma}_{c}=  \widetilde{G}_{0c}^{-1} - \widetilde{G}_{c}^{-1}   \, \,\,\,\,\,\,\,\,
           \widetilde{\Gamma}_{c}=  \widetilde{\Pi}_{0c}^{-1}  + \alpha \widetilde{\Pi}_{c}^{-1} ,
                    \label{133}
\end{eqnarray}
where $G_{0}^{-1}$ and $\Pi_{0}^{-1}$ are given by Eqs. (5) and (6), respectively.

At this stage, one could apply the HS transformation to Eq. (131) to introduce phonon fields.
This is not a change to the EVCA equations but a possible way to solve the reference system.
However, there is another essentially different approach to construct EVCA itself.
One can apply the HS transformation to the original electron action Eq. (4), obtain an
equivalent electron-phonon action, and construct the EVCA for the nonlocal phonon-phonon interaction term.
In this approach, after the Hartree contribution to the phonon term is singled out, we introduce
the source fields for the electron and phonon operators as before, and obtain the action
\begin{eqnarray}
  & & \widetilde{S}_{B} [ c^{*}, c, \widetilde{G}_{0}, \widetilde{Q}_{0} ]        \nonumber \\
   &=&  \int_{0}^{\beta} d \tau \int_{0}^{\beta} d \tau^{\prime} \sum_{i j \sigma} c_{i \sigma}^{*} \left(\tau \right)
          [ - \widetilde{G}_{0}^{-1} ]_{i j \sigma} \left( \tau - \tau^{\prime} \right) c_{j \sigma} \left( \tau^{\prime} \right)          \nonumber \\
   &+&   \int_{0}^{\beta} d \tau \int_{0}^{\beta} d \tau^{\prime} \sum_{i j} : \phi_{i} \left(\tau \right) :
          [ - \widetilde{Q}_{0}^{-1} ]_{i j} \left( \tau - \tau^{\prime} \right) : \phi_{j} \left( \tau^{\prime} \right) :           \nonumber \\
   &+& i \int_{0}^{\beta} d \tau \sum_{i \sigma} \phi_{i}c_{i \sigma}^{*} \left(\tau \right)c_{i \sigma} \left( \tau \right)  \nonumber \\
   &+&  \int_{0}^{\beta} H^{\prime}_{loc B} \left(\tau \right) d \tau                      ,          \label{134}
\end{eqnarray}
where
\begin{eqnarray}
H^{\prime}_{loc B}\left( \tau \right) &=&
H_{loc}\left( \tau \right)+ 2\sum_{ij} \phi_{i}\left( \tau \right) \left[ -Q_{0}^{-1}\right]_{ij} \langle \phi_{j} \rangle \nonumber \\
                &+& \sum_{ij} Q_{0 ij}^{-1} \langle \phi_{i} \rangle \langle \phi_{j} \rangle         \label{135}
\end{eqnarray}
and
\begin{equation}
  Q_{0}^{-1}=\frac{1}{4} \Pi_{0}    .   \label{136}
\end{equation}
The actions $\widetilde{S}_{A}$ and $\widetilde{S}_{B}$ are equivalent in the
sense that for $ \widetilde{Q}_{0c}^{-1}=1/4 \widetilde{\Pi}_{0c}$, they give the same
thermodynamical average values for the same operator.
$\widetilde{G}$ is defined as in Eq. (13). We define the
density-density Green's function, phonon Green's function, and phonon self-energy as
\begin{eqnarray}
      && \widetilde{\Pi}_{B ij} \left(\tau - \tau^{\prime} \right)
      =  \langle T_{\tau} \left[ n_{i} \left(\tau \right) -\langle n_{i} \rangle \right]
                 \left[ n_{j } \left(\tau^{\prime} \right) - \langle n_{j} \rangle \right] \rangle_{\widetilde{S}_{B}}  , \nonumber\\
    && \widetilde{Q}_{ij} \left(\tau - \tau^{\prime} \right)
      =  \langle T_{\tau} \left[ \phi_{i} \left(\tau \right) -\langle \phi_{i} \rangle \right]
                 \left[ \phi_{j } \left(\tau^{\prime} \right) - \langle \phi_{j} \rangle \right] \rangle_{\widetilde{S}_{B}}  , \nonumber\\
   && \widetilde{T} = \widetilde{Q}_{0}^{-1}+ \alpha \widetilde{Q}^{-1}   .      \label{137}
\end{eqnarray}
For a system defined by the action $\widetilde{S}_{B}$, there exist the following equations connecting the
electron and phonon quantities:\cite{Sun1}
\begin{eqnarray}
 && \langle n_{i} \rangle = - 2 i \sum_{j} Q_{0 ij}^{-1} \langle \phi_j \rangle ,  \nonumber \\
 && \widetilde{Q}=-\frac{1}{2} \widetilde{Q}_{0}- \frac{1}{4}\widetilde{Q}_{0}\widetilde{\Pi}_{B}\widetilde{Q}_{0} .
   \label{138}
\end{eqnarray}

Following the previous procedure, the EVCA equations for action $\widetilde{S}_{B}$ are readily obtained as
\begin{eqnarray}
  &&   \widetilde{G}_{c} = \left|  \left( G_{0}^{-1} - \widetilde{\Sigma}_{c}  \right)^{-1} \right|  , \nonumber \\
  &&   \widetilde{Q}_{c} = - \alpha \left|  \left( Q_{0}^{-1} - \widetilde{T_{c}}  \right)^{-1} \right| , \nonumber \\
  &&  \widetilde{\Sigma}_{c}=  \widetilde{G}_{0c}^{-1} - \widetilde{G}_{c}^{-1} ,  \, \,\,\,\,\,\,\,\,
           \widetilde{T}_{c}= \widetilde{Q}_{0c}^{-1}  + \alpha  \widetilde{Q}_{c}^{-1}    .       \label{139}
\end{eqnarray}
It is seen that the two set of equations Eqs. (131) (133) and Eqs. (134) (139) are similar,
but the operators appearing in the corresponding reference actions $n_{i}$ and $\phi_{i}$ have different nature.
Besides Eq. (136), the only direct connection between the two sets of equations occurs when
$ \widetilde{Q}_{0c}^{-1}=1/4 \widetilde{\Pi}_{0c}$ is satisfied in the self-consistent calculations.
However, even at this special point where the two actions are equivalent,
the two set of EVCA equations are different.
In such case, considering $\widetilde{\Pi}_{B}=\widetilde{\Pi}$ and Eqs. (136) (138),
we reexpress the phonon equation
 $\widetilde{Q}_{c} = - \alpha |  ( Q_{0}^{-1} - \widetilde{T_{c}}  )^{-1} |$
 by the density-density Green's function $\widetilde{\Pi}$ defined by $\widetilde{S}_{A}$ as
 \begin{equation}
 \left( \widetilde{\Pi}_{0c}- \widetilde{\Gamma}_{c}^{-1} \right)^{-1} = \left| \left( \Pi_{0} -
                     \widetilde{\Gamma}_{c}^{-1}\right)^{-1} \right|  .   \label{140}
\end{equation}
This equation is different from the corresponding one in Eq. (133). In general,
$\widetilde{Q}_{0c}^{-1}=1/4 \widetilde{\Pi}_{0c}$ is not guaranteed in the
EVCA self-consistent calculations, and we have
even less connection between the two approaches. Therefore our conclusion is
that in general there is no reason for the two set of EVCA equations to give
 out the same results. This is true also for discrete EVCA formalism or for EDMFT, the single-impurity limit
 of the continuous EVCA. We should therefore take into account this difference when we
 compare the results from different EDMFT or EVCA calculations.

It is noted that in the HS transformation approach, besides the electron field we deal with real phonon fields.
This seems to provide a justification of the selection of $\alpha=1/2$ since
it leads to the exact solution $\widetilde{Q}=-1/2 \widetilde{Q}_{0}$, $\widetilde{G}=\widetilde{G}_{0}$
 in the limit of the nonelectron-phonon interaction and $U=0$.
 However, since the electron-phonon interaction
is a fixed finite value in the reference system, this limit is actually not reachable in the EVCA calculations.
A clear criterion for the selection of the $\alpha$ value is therefore still lacking in the present theory.

\subsection{EVCA for correlated spin systems and boson systems}
Our constructions of EVCA in Secs. II and III have the interesting advantage of universality.
It is free to introduce different source fields for each pair of nonlocally coupled
operators, no matter it is a pair of one-particle operators such as $c_{i}$ (for the usual VCA),
or a many-particle one such as $n_{i}$ (for density-density interaction) or
$c_{i\sigma} n_{i \overline{\sigma}}$ (for correlated hopping).
Once the source fields are introduced for these operators, the subsequent derivation procedures are the same,
and similar EVCA equations are obtained. Except in the definitions of the
Green's functions, the nature of the nonlocally coupled operators
does not play a role in the derivation. Therefore, it is rather straightforward to apply these derivations
to systems other than correlated fermions. At the same time, the Luttinger-Ward functional that is most frequently
utilized in correlated fermion systems, can now be generalized for correlated spin systems and boson systems.
This also holds for the EDMFT. For example, the EDMFT equations for the Kondo lattice
model with RKKY interactions can reduce to that for the Heisenberg model by simply discarding the electron degrees
of freedom. As shown in the following, the EDMFT for Heisenberg model will be recovered by our EVCA
in the continuous bath limit. We also discuss the EVCA for correlated boson systems.

We start from the Heisenberg model
\begin{equation}
     H=-\sum_{ \langle i j \rangle} J_{ij}{\bf S}_{i} {\bf} \cdot {\bf S}_{j}    ,  \label{141}
\end{equation}
where $\langle ij \rangle$ denotes summation over pairs of sites. We introduce the
source field $\widetilde{\Pi}_{0}^{-1}$
to each component of the spin operators. The resulting action reads

\begin{eqnarray}
   && S \left[ {\bf S}, \widetilde{\Pi}_{0} \right] = S_{b}  \nonumber \\
      && + \int_{0}^{\beta} d \tau \int_{0}^{\beta} d \tau^{\prime} \sum_{i,j, \eta} S_{i}^{\eta} \left(\tau \right)
           \left[ -\widetilde{\Pi}_{0}^{-1}  \right]_{ij}^{\eta} \left( \tau- \tau^{\prime}
	   \right) S_{j}^{\eta} \left(\tau^{\prime} \right)    .  \nonumber \\
	 &&   \label{142}
\end{eqnarray}
Here, $S_{b}$ is the contribution from Berry's phase. The original system is realized when
\begin{eqnarray}
   \left[ \widetilde{\Pi}_{0}^{-1}  \right]_{ij}^{\eta} \left( \tau- \tau^{\prime} \right)&=&
         \left[ \Pi_{0}^{-1}  \right]_{ij}^{\eta} \left( \tau- \tau^{\prime}  \right)       \nonumber \\
	&=& \frac{1}{2} J_{ij} \delta \left ( \tau -\tau^{\prime} \right)     .   \label{143}
\end{eqnarray}
 The partition function $\widetilde{Z}$ and the free energy $\widetilde{F}$ are
    \begin{eqnarray}
       && \widetilde{Z} = \int \prod_{i} {\mathcal D} {\bf S}_{i} e^{-S \left[ {\bf S}, \widetilde{\Pi}_{0} \right] } ,    \nonumber \\
       && \widetilde{F} = -\frac{1}{\beta} \ln \widetilde{Z}    .   \label{144}
   \end{eqnarray}
We introduce the Green's functions
\begin{equation}
  \widetilde{\Pi}_{ij}^{\eta}\left( \tau- \tau^{\prime}  \right) = \frac{1}{\widetilde{Z}} \int \prod_{i} {\mathcal D} {\bf S}_{i}
                     \left[ S_{i}^{\eta}\left( \tau \right)S_{j}^{\eta} \left( \tau^{\prime} \right)
		    e^{-S \left[ {\bf S}, \widetilde{\Pi}_{0} \right] }   \right]    \label{145}
\end{equation}
and the self-energy
\begin{equation}
      \widetilde{ \Gamma } =\widetilde{\Pi}_{0}^{-1} + \alpha \widetilde{\Pi}^{-1}  .   \label{146}
\end{equation}
Here, $\eta =x, y, z$ are the spin components. $\alpha=1/2$ is usually used. The
latter equation is regarded as the matrix equation in the lattice, time, and spin space.
The same procedure as in Sec. II will lead to the following form of the free energy functional:
\begin{equation}
   \widetilde{F}_{LW} \left[ \widetilde{\Pi} \right]= \frac{1}{\beta} \widetilde{\Phi}\left[ \widetilde{\Pi} \right]
   + \frac{1}{\beta} {\text Tr} \ln \widetilde{\Pi} -\frac{1}{\beta} {\text Tr} \left[ \Pi_{0}^{-1} \widetilde{\Pi} \right]  .  \label{147}
\end{equation}
Here, the trace is carried out in the lattice coordinate, time, and spin space.
$\widetilde{F}_{LW}$ is stationary at the physical value of the Green's function, and up to a constant,
its value at the stationary point is the physical value of the free energy.
The derivative of the generalized Luttinger-Ward functional
$\Phi[ \widetilde{\Pi} ]$ with respect to $\widetilde{\Pi}$ gives the self-energy.
The free-energy functional necessary for the construction of EVCA equation reads
\begin{eqnarray}
&&     \widetilde{F}_{EVCA}\left[\widetilde{\Gamma} \right]  \nonumber \\
  &=& \widetilde{F}\left[\widetilde{\Gamma} \right]
    +\frac{1}{\beta} {\text Tr} \ln \left( \widetilde{\Pi}_{0}^{-1} - \widetilde{\Gamma}\right)
      -\frac{1}{\beta} {\text Tr} \ln \left( {\Pi}_{0}^{-1} -\widetilde{\Gamma}\right)   . \nonumber\\
    &&   \label{148}
\end{eqnarray}
From this functional, the EVCA equations for the Heisenberg model are obtained as
\begin{eqnarray}
  &&   \widetilde{\Pi}_{c} = -\alpha \left| \left( {\Pi}_{0}^{-1} - \widetilde{\Gamma}_{c}  \right)^{-1} \right| , \nonumber \\
  &&  \widetilde{\Gamma}_{c}= \widetilde{\Pi}_{0c}^{-1} + \alpha \widetilde{\Pi}_{c}^{-1}
              .  \label{149}
\end{eqnarray}
Equation (149) is formally the same as the general VCA equation for electrons, but with the reference system defined as
\begin{eqnarray}
    && S_{ref} \left[ {\bf S}, \widetilde{\Pi}_{0c} \right]    \nonumber \\
    =&& \int_{0}^{\beta} d \tau \int_{0}^{\beta} d \tau^{\prime} \sum_{ij \alpha}   \nonumber \\
     & & S_{i}^{\alpha} \left( \tau \right) \left[-\widetilde{\Pi}_{0 c}^{-1} \right]_{ij}^{\alpha} \left( \tau - \tau^{\prime}\right)
         S_{j}^{\alpha} \left( \tau^{\prime} \right) .   \label{150}
\end{eqnarray}
For a discrete reference system, $\widetilde{ \Pi}_{0c}^{-1}$ can be parametrized by some parameters $\widetilde{J}$,
and the first equation in Eq. (149) is replaced with
\begin{equation}
    {\text Tr} \left[ \widetilde{\Pi}_{c} +\alpha \left|  \left( {\Pi}_{0}^{-1} - \widetilde{\Gamma}_{c}  \right)^{-1} \right| \right]
            \frac{\delta \widetilde{\Gamma}_{c}\left( \widetilde{t} \right) }{\delta \widetilde{J}}
	     =0         .  \label{151}
\end{equation}
With this equation, it is possible to study the Heisenberg model with discrete reference systems.

The EDMFT has been used to study the coupled fermion-boson system\cite{Motome1} where the bosons
are phonons. Recently the correlated boson systems is studied intensively in the context of cold boson atoms in the
optical lattice. The bosonic Hubbard model\cite{Fisher1} is one of the simplest models to describe the boson atoms
which have particle number conservation.
\begin{equation}
H =  -\sum\limits_{i,j} t_{i j} a_{i }^{\dagger}a_{j } -\mu \sum_{i} a_{i }^{\dagger}a_{i } + H_{loc}   \,\, ,  \label{152}
\end{equation}
where $H_{loc}=U \sum_{i} ( a_{i }^{\dagger}a_{i }-1) a_{i }^{\dagger}a_{i }$
describes the local repulsion between bosons.
The ideal EVCA for this model should be able to describe both the
Bose-Einstein consensation phase and the Mott insulator phase.
For this problem, it is important to know that what kind of Weiss fields are necessary for
 the description of the physics.
 From the usual splitting of the
boson field into sum of condensate wave function and the quasiparticle field, it is plausible that besides
the normal boson Green's function, the new Weiss fields should include the condensate
fields as well as the anomalous Green's
functions.\cite{Kita1} The corresponding self-energy functional and EVCA equation can be
 derived in a similar way as for fermions. Detailed work in this direction is in progress.
It would be interesting to explore the EVCA for correlated bosons
and make comparison with the results from other approaches.

\subsection{EVCA for classical systems}
In this part, we derive the EVCA equations for a classical system, the ferromagnetic Ising model.
The purpose is to show that EVCA applied on larger clusters may be used as a useful tool for the study
of classical systems.
Here we consider the classical Ising model
\begin{equation}
     H= - \sum_{\langle ij \rangle}J_{ij} S_{i}^{z} S_{j}^{z}   .  \label{153}
\end{equation}
$S_{i}^{z}$ is the $z$ component of spin $1/2$. The summation is over pairs,
and we assume $J_{ij} > 0$. Because all the operators present in this model commute with
each other, there is no quantum fluctuations in this model.
In this case, instead of introducing a time-dependent source field as for the Heisenberg model, it is sufficient to
introduce static ones $[ \widetilde{\Pi}_{oc}^{-1} ]_{ij}$. Taking into account of the possible symmetry-broken phase,
the reference system with an open cluster boundary condition has the action
\begin{eqnarray}
    && S_{ref} \left[ S^{z}, \widetilde{\Pi}_{0c} \right]    \nonumber \\
    =&& \beta \sum_{ij} \left( S_{i}^{z} -m_{i} \right) \left[-\widetilde{\Pi}_{0 c}^{-1} \right]_{ij} \left(S_{j}^{z} - m_{j} \right)
    \nonumber \\
      &-& \beta\sum_{\langle ij \rangle} J_{ij} \left( S_{i}^{z} m_{j} + S_{j}^{z} m_{i} - m_{i} m_{j} \right)    , \label{154}
\end{eqnarray}
and the partition function
\begin{equation}
   \widetilde{Z} = {\text Tr} e^{-S_{ref} \left[ \widetilde{\Pi}_{0c}\right]}  .   \label{155}
\end{equation}
Following the same procedure as before, we get the following EVCA equations:
\begin{eqnarray}
   &&\widetilde{\Pi}_{c ij}= \beta \langle \left( S_{i}^{a} - m_{i} \right)  \left( S_{j}^{a} - m_{j} \right) \rangle_{S_{ref}},  \nonumber \\
   &&\widetilde{\Gamma}_{c} = \widetilde{\Pi}_{0c}^{-1} + \alpha \widetilde{\Pi}_{c}^{-1}  ,   \nonumber \\
   &&\widetilde{\Pi}_{c} =-\alpha \left| \left( \Pi_{0}^{-1}- \widetilde{\Gamma}_{c} \right)^{-1} \right|  ,  \nonumber \\
   && m_{i}= \langle S_{i}^{z}\rangle_{S_{ref}} , \, \, \,\,\,\,  \Pi_{0 ij}^{-1}=\frac{1}{2} J_{ij} .   \label{156}
\end{eqnarray}
Equations (154)-(156) form a closed set of self-consistent equations for the Ising model. The Weiss fields are real variables.

For the simplest case of the one-site cluster, we obtain the following self-consistent equations:
\begin{eqnarray}
  && \beta \left(\frac{1}{4} - m^{2} \right)    \nonumber \\
  &=& -\frac{1}{2N} \sum_{k} \frac{1}{-1/2 J_{k}- \widetilde{\Pi}_{0}^{-1}
                           -1/[2\beta \left(1/4-m^{2} \right)]   }   \nonumber \\
  &&  m=-\frac{1}{2} \tanh\left[\beta \left( \widetilde{\Pi}_{0}^{-1} +\frac{1}{2} J_0 \right)m \right]  .  \label{157}
  \end{eqnarray}
This set of equations is different from the usual Weiss mean-field equations.
It has been obtained (with different parameter definitions) by Pankov {\it et al.}
in a semiclassical analysis of EDMFT.\cite{Pankov1} It was found that this mean-field theory predicts a
better phase transition temperature than the Weiss mean-field theory. However, the order of
the transition is given incorrectly. Instead of a second-order transition, this one-site theory gives a first-order
phase transition for dimensions $d < 4$. It is expected that the approximation will be improved by
 using a larger cluster size. However, it is still an open question whether the correct order of the phase transition
 can be obtained in EVCA with larger clusters.

\section{Summary}

In this paper, we extend the VCA to include the nonlocal interactions. A universal way of constructing
this theory is described for different systems, i.e., the
Hubbard model with density-density interaction, the Hubbard model with correlated hopping, the Heisenberg
model, and the Ising model. The construction method proposed here is universal, and may deserve further
consideration in the context of the general variational principle. \cite{Potthoff4} Taking the extended Hubbard model as an example,
we show that the simplest three-site EVCA can already produce results very close to those of the EDMFT.
The latter can only be realized in the limit of a continuous reference system.
A number of existing theories are obtained as limiting cases of EVCA.
We arrive at a cluster extension of EDMFT for a system with nonlocal interactions,
and arrive at a cluster extension of the DMFTCH for systems with correlated hopping.
VCA (EVCA)
equations which are based on clusters with periodic boundary conditions are also proposed.
This produces translation-invariant solutions and has the DCA (DCA extended to include nonlocal
interaction) as the continuous limit. The high efficiency and flexibility of the EVCA makes it possible
to study a large class of correlated systems with less numerical cost
by adopting a relatively simple reference system. Some interesting issues about EVCA are discussed.

\section{Acknowledgments}

The author is grateful to the helpful discussions with Matthias Vojta. This work is supported
 by the DFG through the Graduiertenkolleg GRK 284.


\begin{thebibliography}{99}


 \bibitem{Metzner1} W. Metzner and D. Vollhardt, Phys. Rev. Lett. {\bf 62}, 324 (1989).
 \bibitem{Georges1} A. Georges, G. Kotliar, W. Krauth, and M. J. Rozenberg, Rev. Mod. Phys. {\bf 68}, 13 (1996).
 \bibitem{Bulla1} R. Bulla, Phys. Rev. Lett. {\bf 83}, 136 (1999).
 \bibitem{Tong1} N. H. Tong, S. Q. Shen, and F. C. Pu, Phys. Rev. B {\bf 64}, 235109 (2001).
 \bibitem{Wahle1} J. Wahle, N. Bl\"{u}mer, J. Schlipf, K. Held, and D. Vollhardt, Phys. Rev. B {\bf 58}, 12749 (1998).
 \bibitem{Vollhardt1} D. Vollhardt, N. Bl\"{u}mer, K. Held, M. Kollar, J. Schlipf, M. Ulmke, and J. Wahle,
                                Adv. Solid State Phys. {\bf 38}, 383 (1999).
\bibitem{Millis1} A. J. Millis, Boris I. Shraiman, and R. Mueller, Phys. Rev. Lett. {\bf 77}, 175 (1996).
 \bibitem{Held1} K. Held, I.A. Nekrasov, G. Keller, V. Eyert, N. Blümer, A.K. McMahan, R.T. Scalettar,
                          T. Pruschke, V.I. Anisimov, and D. Vollhardt, in {\it Quantum Simulations of Complex Many-Body
			  Systems: From Theory to Algorithms},  edited by J. Grotendorst, D. Marks, and A. Muramatsu,
			  NIC Series Vol. 10 (NIC directors, Forschungszentrum Julich, 2002), pp. 175-209.
\bibitem{Vollhardt2} D. Vollhardt, K. Held, G. Keller, R. Bulla, Th. Pruschke, and I.A. Nekrasov, V.I. Anisimov,
               in {\it Kondo Effect -- 40 Years after the Discovery },
	       special issue of J. Phys. Soc. Jpn. {\bf 74}, 136 (2005).
\bibitem{Schiller1} A. Schiller and K. Ingersent, Phys. Rev. Lett. {\bf 75}, 113 (1995).
\bibitem{Zarand1} G. Zarand, D. L. Cox, and A. Schiller, cond-mat/0002073 (unpublished).
 \bibitem{Kotliar1} G. Kotliar, S. Y. Savrasov, G. Palsson, and G. Biroli, Phys. Rev. Lett. {\bf 87}, 186401 (2001).
 \bibitem{Biroli1} G. Biroli and G. Kotliar, Phys. Rev. B {\bf 65}, 155112 (2002).
 \bibitem{Biroli2} G. Biroli, O. Parcollet, and G. Kotliar, Phys. Rev. B {\bf 69}, 205108 (2004).
 \bibitem{Bolech1} C. J. Bolech, S. S. Kancharla, and G. Kotliar, Phys. Rev. B {\bf 67}, 075110 (2003).
 \bibitem{Hettler1} M. H. Hettler, A. N. Tahvildar-Zadeh, M. Jarrell, T. Pruschke, and H. R. Krishnamurthy, Phys. Rev.
                             B {\bf 58}, R7475 (1998);
                             M. H. Hettler, M. Mukherjee, M. Jarrell, and H. R. Krishnamurthy, {\it ibid.} {\bf 61}, 12739 (2000);
	                      Th. A. Maier, M. Jarrell, A. Macridin, and C. Slezak, Phys. Rev. Lett. {\bf 92}, 027005 (2004),
			      and references therein.
\bibitem{Si1} Q. Si and J. L. Smith, Phys. Rev. Lett. {\bf 77}, 3391 (1996);
                       J. L. Smith and Q. Si, Europhys. Lett. {\bf 45}, 228 (1999).
\bibitem{Kajueter1} H. Kajueter, Ph.D. thesis, Rutgers University, 1996.
\bibitem{Sun1} P. Sun and G. Kotliar, Phys. Rev. B {\bf 66}, 085120 (2002), and references therein.
\bibitem{Schiller2} A. Schiller, Phys. Rev. B {\bf 60}, 15660 (1999).
\bibitem{Shvaika1} A. M. Shvaika, Phys. Status. Solidi B {\bf 236}, 368 (2003); Phys. Rev. B {\bf 67}, 075101 (2003).
\bibitem{Okamoto1} S. Okamoto, A. J. Millis, H. Monien, and A. Fuhrmann, Phys. Rev. B {\bf 68}, 195121 (2003).
\bibitem{Maier1} T. Maier, M. Jarrell, Th. Pruschke, and M. H. Hettler, cond-mat/0404055 (unpublished).
\bibitem{Potthoff1} M. Potthoff, Euro. Phys. J. B {\bf 32}, 429 (2003); {\bf 36}, 335 (2003); M. Potthoff, M. Aichhorn,
                              and C. Dahnken, Phys. Rev. Lett. {\bf 91}, 206402 (2003).
\bibitem{Potthoff2} M. Potthoff, Phys. Rev. B {\bf 64}, 165114 (2001).
\bibitem{Potthoff3} M. Potthoff, cond-mat/0406671 (unpublished).
\bibitem{Luttinger1} J. M. Luttinger and J. C. Ward, Phys. Rev. {\bf 118}, 1417 (1960).
\bibitem{Baym1} G. Baym, Phys. Rev. {\bf 127}, 1391 (1962).
\bibitem{Baym2} G. Baym and L. P. Kadanoff, Phys. Rev. {\bf 124}, 287 (1961).
\bibitem{note1} Our definition of $\Phi$ follows that of Baym (Ref. 28), and is different by a factor of
                            $\beta$ from that of Refs. 24 and 27.
\bibitem{Dahnken1} C. Dahnken, M. Aichhorn, W. Hanke, E. Arrigoni, and M. Potthoff, Phys. Rev. B {\bf 70}, 245110 (2004).
\bibitem{Aichhorn1} M. Aichhorn, H. G. Evertz, W. von der Linden, and M. Potthoff, Phys. Rev. B {\bf 70}, 235107 (2004).
\bibitem{note3} For examples, in Refs. 19 and 38, the selection of $\alpha=1/2$ is implied by the coefficient $1/2$ in
                        the action or Hamiltonian and by the usual definition of the self-energies.
\bibitem{Chitra2} R. Chitra and G. Kotliar, Phys. Rev. Lett. {\bf 84}, 3678 (2000).
\bibitem{Chitra1} R. Chitra and G. Kotliar, Phys. Rev. B {\bf 63}, 115110 (2001).
\bibitem{Fukuda1} R. Fukuda, M. Komachiya, S. Yokojima, Y. Suzuki, K. Okumura, and T. Inagaki,
                              Prog. Theor. Phys. Suppl. {\bf 121}, 1 (1995).
\bibitem{note2} Up to a constant.
\bibitem{Smith1} J. L. Smith and Q. Si, Phys. Rev. B {\bf 61}, 5184 (2000).
\bibitem{Pozgajcic1} K. Pozgajcic, cond-mat/0407172 (unpublished).
\bibitem{note4} The coefficient $e^{i \omega_n 0^{+}}$ produces a term $\int d \epsilon \, \epsilon \rho_0 (\epsilon)$,
                         which is zero for symmetric $\rho_0(\epsilon)$.
\bibitem{Bulla2} R. Bulla, H.-J. Lee, N.-H. Tong, and M. Vojta, Phys. Rev. B {\bf 71}, 045122 (2005).
\bibitem{Kollar1} M. Kollar and D. Vollhardt, Phys. Rev. B {\bf 63}, 045107 (2001).
\bibitem{Aryanpour1} K. Aryanpour, Th. A. Maier, and M. Jarrell, Phys. Rev. B {\bf 71}, 037101 (2005);
                                 G. Biroli and G. Kotliar, {\it ibid.} {\bf 71}, 037102 (2005).
\bibitem{Motome1} Y. Motome and G. Kotliar, Phys. Rev. B {\bf 62}, 12800 (2000).
\bibitem{Fisher1} M. P. A. Fisher, P. B. Weichman, G. Grinstein, and D. S. Fisher, Phys.
                            Rev. B {\bf 40}, 546 (1989).
\bibitem{Kita1} Takafumi Kita, J. Phys. Soc. Jpn. {\bf 74}, 1891 (2005).
\bibitem{Pankov1} Sergey Pankov, Gabriel Kotliar, and Yukitoshi Motome, Phys. Rev. B {\bf 66}, 045117 (2002).
\bibitem{Potthoff4} M. Potthoff, cond-mat/0503715 (unpublished).
\bibitem{Maier2} Th. A. Maier, cond-mat/0312447 (unpublished); Physica B {\bf 359-361}, 512 (2005).
\end{thebibliography}
\end{document}